\documentclass{aastex63}

\usepackage{lineno}
\usepackage{amsmath}
\usepackage{array}
\usepackage{graphicx}
\usepackage{algorithm}
\usepackage{algorithmic}

\shorttitle{Gamma-Rray Radiation of the Crab Nebula}
\shortauthors{Nie et al.}
\begin{document}

\title{Ultrahigh-energy Gamma-Ray Radiation from the Crab Pulsar Wind Nebula}

\correspondingauthor{Ze-Jun Jiang}
\email{zjjiang@ynu.edu.cn}

\author{Lin Nie}
\affiliation{Department of Astronomy, Yunnan University, and Key Laboratory of Astroparticle Physics of Yunnan Province, Kunming, 650091, People's Republic of China.}
\author{Yang Liu}
\affiliation{Department of Astronomy, Yunnan University, and Key Laboratory of Astroparticle Physics of Yunnan Province, Kunming, 650091, People's Republic of China.}
\author{Zejun Jiang}
\affiliation{Department of Astronomy, Yunnan University, and Key Laboratory of Astroparticle Physics of Yunnan Province, Kunming, 650091, People's Republic of China.}
\author{Xiongfei Geng}
\affiliation{School of Electrical and Information Technology, Yunnan Minzu University, Kunming 650091, Yunnan, People's Republic of China.}

\begin{abstract}

 It has been long debated whether the high-energy gamma-ray radiation from the Crab nebula stems from leptonic or hadronic processes. In this work, we investigate the multiband nonthermal radiation from the Crab pulsar wind nebula with the leptonic and leptonic$-$hadronic hybrid models, respectively. Then we use the  Markov Chain Monte Carlo sampling technology and method of sampling trace to study the stability and reasonability of the model parameters according to the recently observed results and obtain the best-fitting values of parameters. Finally, we calculate different radiative components generated by the electrons and protons in the Crab nebula.
 The modeling results indicate that the pure leptonic origin model with the one-zone only can partly agree with some segments of the data from various experiments (including the PeV gamma-ray emission reported by the LHAASO and the other radiation ranging from the radio to very-high-energy gamma-ray wave band), and the contribution of hadronic interaction is hardly constrained. However, we find that the hadronic process may also contribute, especially in the energy range exceeding the $\rm PeV$. In addition, it can be inferred that the higher energy signals from the Crab nebula could be observed in the future.

\end{abstract}

\keywords{Gamma-rays(637); Radiative process(2055); Rotation powered pulsars(1408); Nebulae(1095)}


\section{Introduction} \label{sec:intro}
It is believed that pulsar wind nebulae (PWNe) are an important high-energy radiative sources in the Galaxy. The PWNe are also known as candidates for Galactic cosmic ray sources. It is suggested that the PWNe inside supernova remnants (SNRs) can further accelerate the relativistic protons accelerated by the SNR shock to the energy of $\rm PeV$ \citep{2018MNRAS.478..926O}. Therefore, as a PeVatron, the PWNe will accelerate the charged particles (electrons and protons) to relativistic energy and produce the multiband nonthermal photons with energies ranging from radio to ultra-high-energy gamma-ray bands.

The Crab Nebula is one of the comprehensively studied celestial objects, and is an important cosmic laboratory for exploration of the nonthermal relativistic processes in astrophysical settings \citep{1995APh.....3..275A}. It has an age of $\rm \thicksim 940~yr$, lies at a distance of $\rm d\thicksim2~kpc$, and is powered by the pulsar PSR J0534+2200 which has a rotational period of $\rm 33.4~ms$ and a period derivative of $\rm 4.23\times10^{-13}~s~s^{-1}$ \citep{1993MNRAS.265.1003L}. It has been observed in radio \citep{1972A&A....17..172B}, FIR \citep{2002A&A...386.1044B}, optical \citep{1993A&A...270..370V}, soft X-rays, gamma rays, and very-high-energy gamma rays \citep{2008ApJ...674.1037A,  2021ChPhC..45b5002A,2001A&A...378..918K, 2019PhRvL.123e1101A}.

\begin{table*}
\tablenum{1}
\begin{center}
\caption{The best-fitting parameters for the leptonic model.}
\footnotesize
\begin{tabular}{@{}ccccccc@{}}
\label{tab:1}\\
\hline\hline
 Model& $ \alpha_{1} $ & $\alpha_{2}$ & $ \eta $ & $B_{PWN}(uG)$ & $E_{cut}(TeV)$ & $\varepsilon_{0}(TeV)$\\
\hline
leptonic& $1.533^{+0.0056}_{-0.0065}$ & $2.51^{+0.0031}_{-0.0033}$& $0.265^{+0.0034}_{-0.0032}$ &$129.42^{+1.8}_{-1.78}$ & $0.154^{+0.0059}_{-0.0061}$& $5.649^{+0.052}_{-0.051}\times10^{3}$ \\
\hline\hline
\end{tabular}
\end{center}
\end{table*}

\begin{figure*}[h]
\centering
\includegraphics[width=0.45\textwidth]{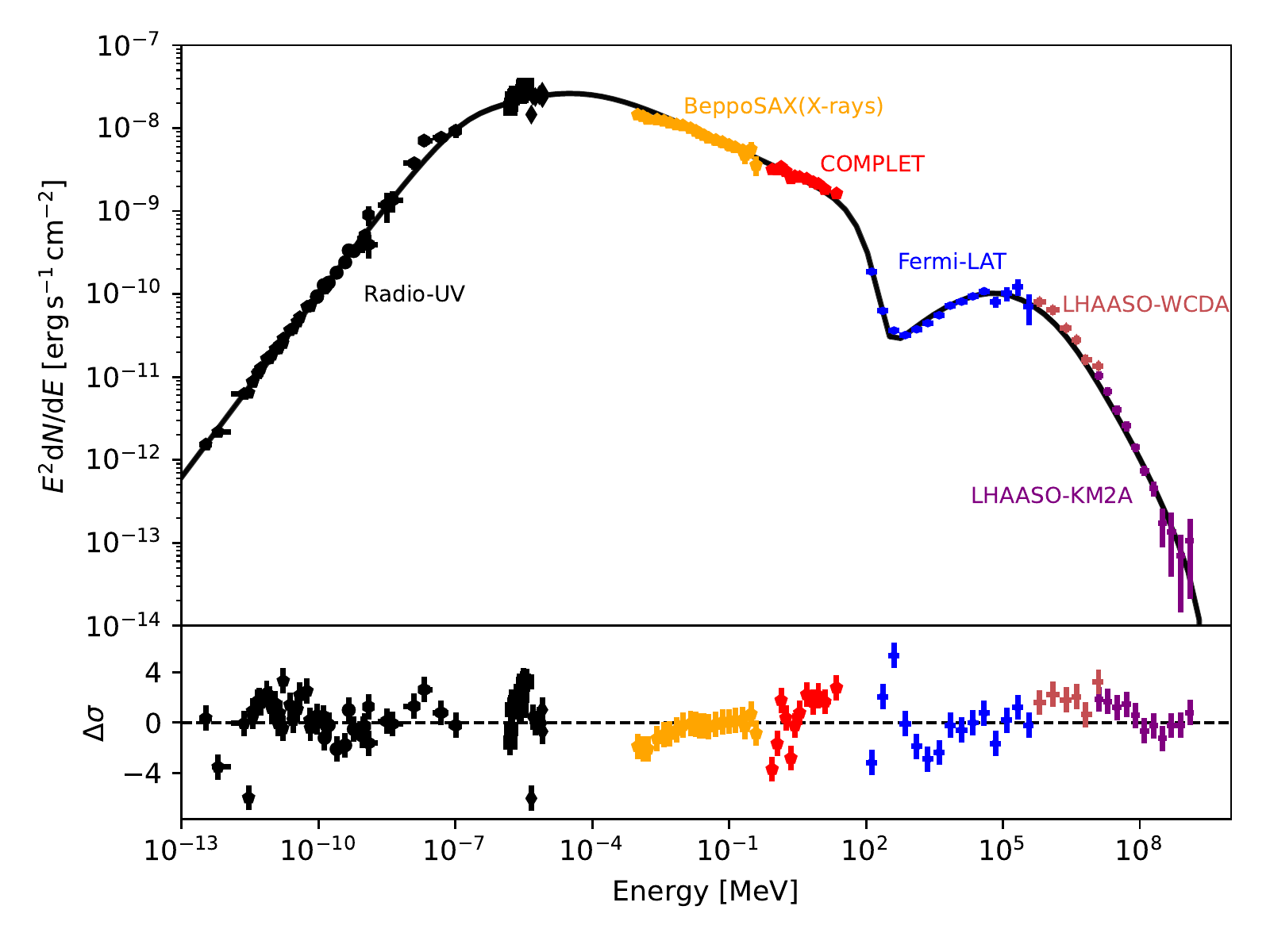}
\includegraphics[width=0.45\textwidth]{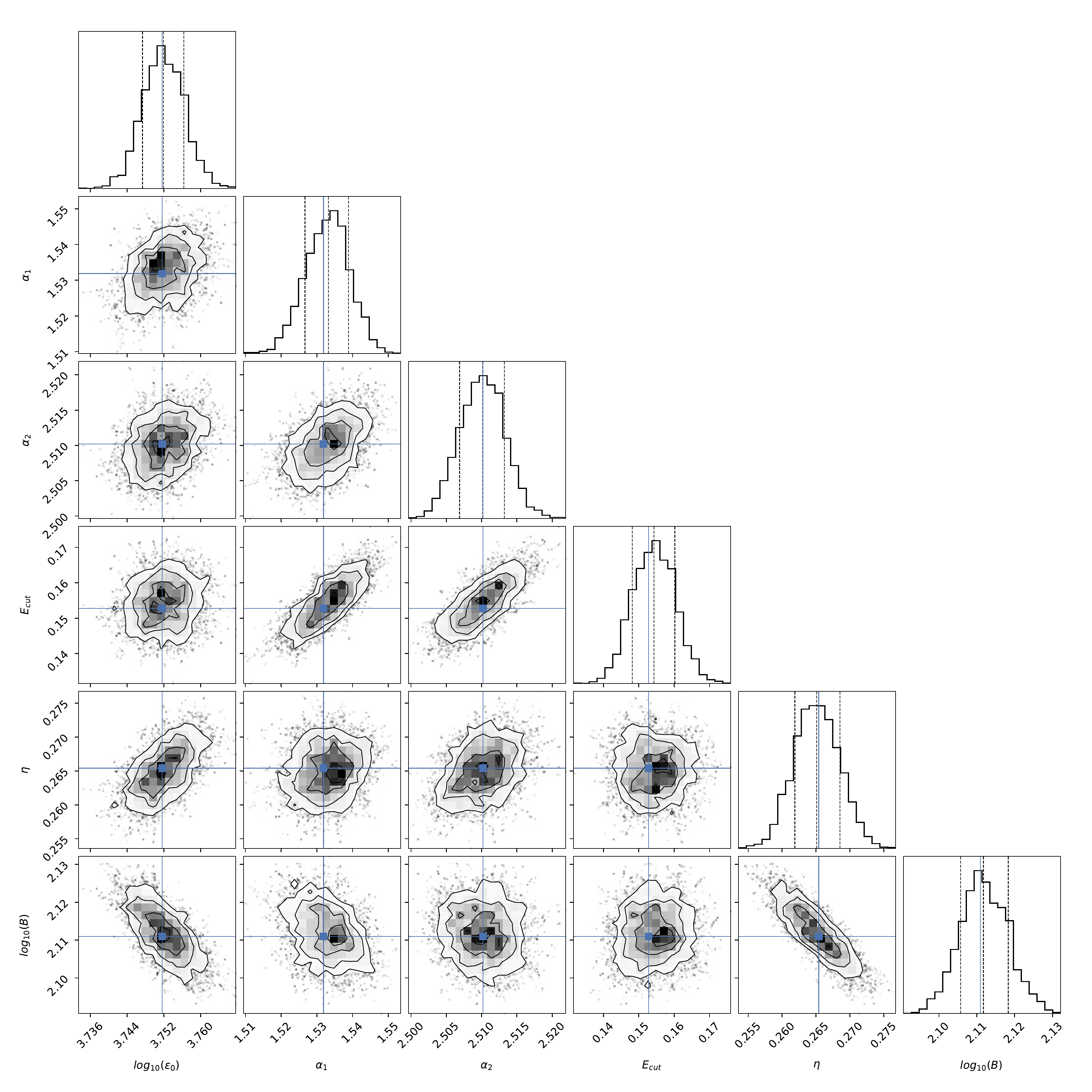}
\caption{The right panel shows the distribution of the model parameters for the one-zone leptonic model, the parameter vector with the highest likelihood found during the run is indicated by the blue cross. The left panel presents the spectrum of the Crab Nebula, which includes the modeling spectrum from the leptonic model, multiwave band  observational data, and the residual of the maximum likelihood model (bottom panel in the left panel plot). The thick black line indicates the maximum likelihood spectrum, and the gray lines present samplings of the posterior distribution of the model parameter vector. The observational data with energy ranging from the radio to UV come from \cite{1972A&A....17..172B}, \cite{2002A&A...386.1044B}, and \cite{1993A&A...270..370V}, the X-ray and MeV data of COMPTEL take from the \cite{2001A&A...378..918K}, the blue data come from the observation of Fermi Large Area Telescope \citep{2020ApJ...897...33A}, the firebrick and purple data points have  been reported by \citep{2021Sci...373..425L}.}
\label{Figure1}
\end{figure*}

\begin{figure*}[h]
\centering
\begin{minipage}{\textwidth}
\includegraphics[width=0.5\textwidth]{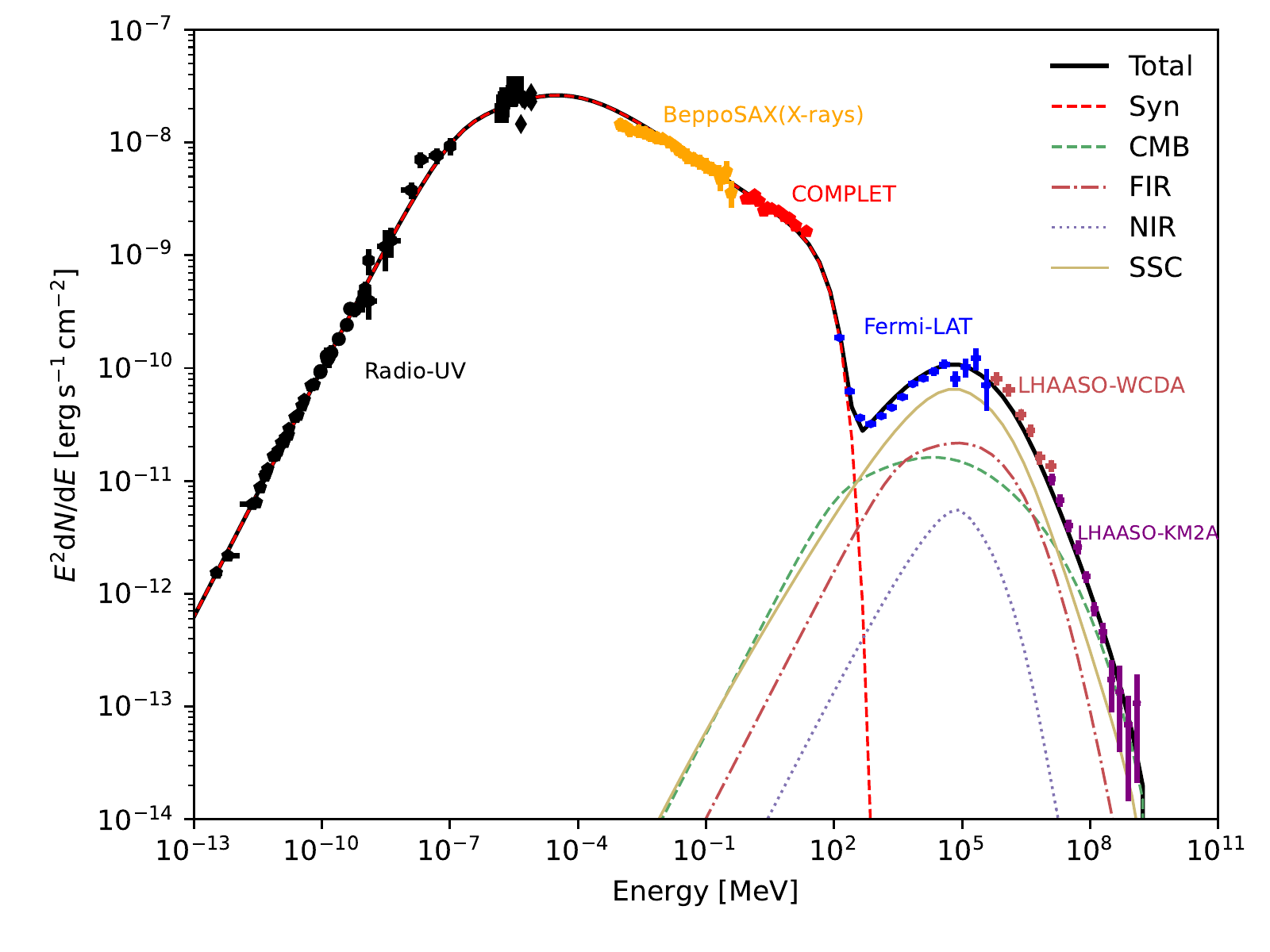}
\includegraphics[width=0.5\textwidth]{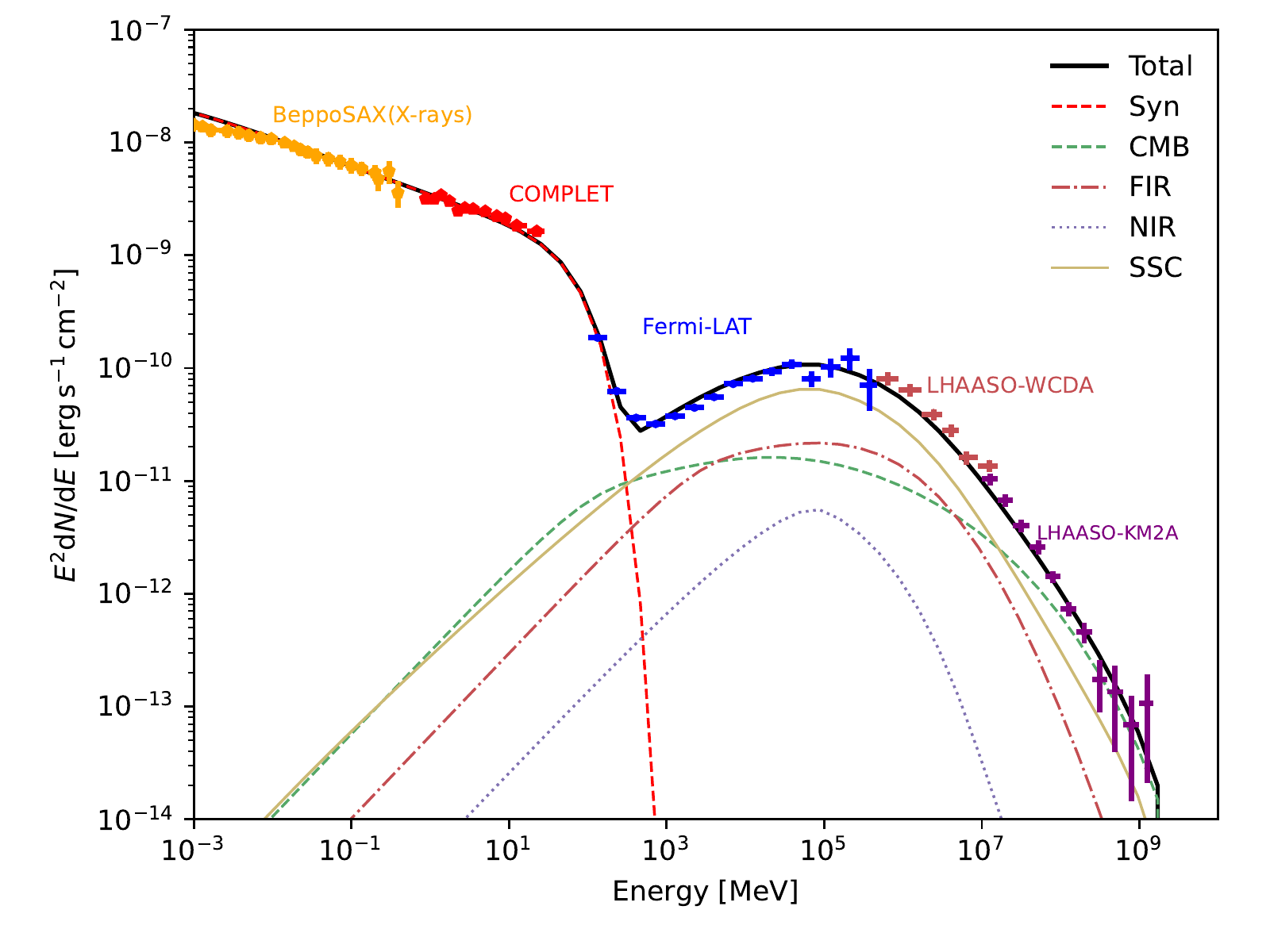}
\end{minipage}
\caption{The Crab nebula SEDs were calculated with the leptonic model. The left panel is a multiwave band nonthermal spectrum and the right panel shows that with energy ranging from X-ray to ultra-high-energy bands. The black solid line is the total spectrum, the red dashed line presents the synchrotron emission, and the IC scatterings with the synchrotron photons (yellow solid line), FIR (dash-doted line), CMB (green dashed line), and NIR (dotted line) are shown. The same as Fig.\ref{Figure1} for the observed data.}
\label{Figure2}
\end{figure*}

It is generally thought that the emission from radio to medium energy gamma rays is generated by the injected electrons through the synchrotron radiative mechanism. The high-energy photon emission mainly comes from Inverse Compton (IC) scattering of the high-energy electrons on the background seed photons, which include the synchrotron background, the cosmic microwave background, and infrared photons in the PWNe  \citep[see, e.g.,][]{2008ApJ...676.1210Z,2010A&A...515A..20F, 2013ApJ...763L...4T,2020MNRAS.498.1911L}. On the other hand, it is also suggested that the gamma rays could be emitted by the hadronic processes. The relativistic protons accelerated in the Crab pulsar outer gap interact with the matter inside the nebula, and this process may contribute in the high-energy gamma-ray range \citep[see, e.g.,][]{1990JPhG...16.1115C,1997PhRvL..79.2616B,2020MNRAS.491.3217K,2021Sci...373..425L}. Therefore, it has been long debated whether the high-energy emission from the PWNe is the leptonic or hadronic origin. The details of the high-energy radiation produced by leptonic process have been discussed for the Crab nebula \citep[see, e.g.,][]{2007whsn.conf...40V,2008ApJ...676.1210Z,2012MNRAS.427..415M}, and that of the gamma-ray emission about the hadronic process have been also investigated \citep[see, e.g.,][]{1990JPhG...16.1115C, 1997PhRvL..79.2616B,2003A&A...407....1B, 2007Ap&SS.309..179B}. However, with the establishment of more and more high-energy observatories, some telescopes have possessed the performance of observing photons of exceeding to the $\rm PeV$ from the astronomic objects. An increasing number of observational data has been reported by the different experiments. For example, the \cite{2019PhRvL.123e1101A} reported that the Tibet air shower array with the underground water-Cerenkov-type muon detector array observed the highest energy photons of exceeding $100~\rm TeV$ with a $5.6\sigma$ statistical significance and pointed the measured spectrum with energy extended to the sub-$\rm PeV$ from the Crab nebula have an absence of high-energy cutoff. Recently, more than 530 photons at energies above $100~\rm TeV$ and up to $1.4~\rm PeV$ from the 12 ultra-high-energy gamma-ray sources with a statistical significance greater than seven standard deviations were reported again by LHAASO \citep{2021Natur.594...33C}. Together with the earlier investigations about the leptonic scenario, the radiative spectrum from the leptons has a cutoff around the sub-PeV region \citep[see, e.g.,][]{2008ApJ...676.1210Z,2012MNRAS.427..415M, 2020MNRAS.497.3477Z}. It seems that the other components of gamma rays have a significant contribution.

In this paper, we investigated multiband nonthermal radiation from the Crab pulsar wind nebula with the leptonic and leptonic$-$hadronic hybrid model, respectively. We simulated the broadband nonthermal radiation which includes components of leptonic and hadronic origin. In addition, we derived the best-fitting values and distribution of the model parameters with the method of the Markov Chain Monte Carlo (MCMC) sampling to analyze the reasonability and stability of fitting parameters in the physics and to further constrain the parameters of the model base on the recently observed data.

\section{Model and Result} \label{sec:model and result}
\subsection{Leptonic Origin Model}\label{sec:lepton}
In this model, the pulsar associated with the PWN loses its rotational energy via a pulsar wind composed of magnetic and high-energy particles to power the high-energy physical process inside the nebula \citep{1996MNRAS.278..525A,2010A&A...515A..20F}. The relativistic wind of particles driven by the pulsar is blown into the ambient medium and generates a termination shock wave, which accelerates the electrons to relativistic energy. These relativistic electrons interact with the magnetic field and low-energy background photons (the synchrotron, thermal, FIR, and microwave background radiation), and generate the multiband nonthermal photons with energies ranging from radio to high-energy gamma-ray bands \citep{2008ApJ...676.1210Z, 2010A&A...515A..20F,2017ApJ...834...43L}. According to the review of the leptonic model  \citep[see, e.g.,][]{2008ApJ...676.1210Z, 2007whsn.conf...40V,2013ApJ...763L...4T}, the electrons injected into PWNe are accelerated by the pulsar magnetosphere and the termination shock. Therefore, the relativistic particles injected into the PWNe are also assumed as two different power-law components from the pulsar magnetosphere and shock acceleration, respectively. The injected spectrum of relativistic electrons inside PWNe is described as

\begin{equation}
\label{equation1}
 Q(E_{e},t)=
\begin{cases}
 Q_{0}(t)(E_{e}/E_{\rm cut})^{-\alpha_{1}} & \text{if $\rm E_{e}<E_{cut}$}\\
 Q_{0}(t)(E_{e}/E_{\rm cut})^{-\alpha_{2}} & \text{if $\rm E_{e}\geq E_{cut}$}
\end{cases}
\end{equation}
where the $Q_{0}$ can be determined by the $\int Q(E_{e},t)E_{e}dE_{e}=\eta L(t)$; $\eta$ is the conversion efficiency from spin-down power into electron luminosity. The maximum energy of the electrons was express as $E_{max}(t)\approx \varepsilon_{0}\sqrt{\frac{L(t)}{L_{0}}}$, and $L_{0}$ is initial spin-down power. The electron energy distribution was given by

\begin{equation}
\label{equation2}
\frac{dN(E_{e},T_{age})}{dt}=\int^{T_{age}}_{0}Q(E_{e},t)exp(-\frac{T_{age}-t}{\tau_{eff}})dt
\end{equation}
where the $\tau_{eff}^{-1}=\tau_{esc}(t)^{-1}+\tau_{syn}(t)^{-1}$, $\tau_{esc}(t)$ is the escape timescale, and $\tau_{\rm syn}(t)$ is the lifetime of the relativistic electron of the synchrotron emission loss. The details of temporal evolution about the electron in the PWNe are discussed by the \cite{2008ApJ...676.1210Z}\cite[also see the version of][]{2010A&A...515A..20F,2017ApJ...834...43L}.

After the electron distribution is determined, we calculate multiband nonthermal radiation via the mechanism of the synchrotron radiation and IC scattering. For the calculation of the synchrotron radiation and IC scattering, we use the formalism of \cite{1970RvMP...42..237B}.

In this model, we consider $\alpha_{1}$, $\alpha_{2}$, $\varepsilon_{0}$, $E_{\rm cut}$, $\eta$, and $B_{PWB}$ as the free parameters that determine the distribution of the injected electrons and the radiative spectra. In the calculative process, we use the MCMC sampling method to derive the best-fitting values of model parameters and the distribution of the free parameters. We obtained the values of parameters and show them in Table.\ref{tab:1}. In Fig.\ref{Figure1} and Fig.\ref{Figure2}, we show modeling results that include the distribution of parameters, and results of comparison between the theoretical spectrum with the best-fitting model parameters and observed data. Furthermore, we used the $plot\_chain$ function of $Naima$ \citep{2015ICRC...34..922Z} to plot the diagnostic figure of sampling about the fitting parameters to judge whether the sampling has stabilized around the maximum likelihood parameters, and the results of the sampling are presented in Fig.\ref{Figure6} (see the Appendix.\ref{appendix A}). Besides, the bottom left of each panel in Fig.\ref{Figure6} (see the Appendix.\ref{appendix A}) provides statistics of parameter distribution with the medium value, which has uncertainties based on the 16th and 84th percentiles.

Finally, we use this model with best-fitting parameters to compute the total energy of electrons in Crab nebula  $W_{e}\approx7.299^{+0.0046}_{-0.0044}\times10^{48}\rm erg$.

As shown in Fig.\ref{Figure1}, some data have a strong signal of deviation from the model. In the range from $1~\rm MeV$ to $500~\rm MeV$, the COMPTEL data seem off from the model, with at least $6\sigma$, and Fermi data in this range are off by at least $10\sigma$, with a totally different trend. In the VHE energy range between 1 and $\thicksim100~\rm TeV$, the observed data is lower than the prediction by $4\sigma$. On the other hand, the observational data have a steeper spectrum than the model predictions between $\thicksim50$ and $\thicksim500~\rm TeV$, while that around $\rm PeV$ have an opposite trend. Importantly, for the $\rm PeV$ photons, the acceleration rate of parent electrons at the site of $\rm PeV$ photons was described as \citep{2021Sci...373..425L} $\eta=0.14(B/100uG)(E_{\gamma}/1PeV)^{1.54}$. We obtained magnetic field strength of $B\sim130~\rm uG$ through fitting multiband data with the one-zone leptonic model. This means that the acceleration rate of the parent electron is larger than $21\%$ for the photons with energies up to $1.1~\rm PeV$.
This scenario is inconsistent with the magnetohydrodynamic \citep[MHD; the acceleration rate of the electron in the young supernova remnants is smaller by exceeding 3 orders of magnitude;][]{2001RPPh...64..429M, 2021Sci...373..425L}. Therefore, it is perhaps necessary to introduce a new component to explain the $\rm PeV$ radiation from the Crab nebula.

\subsection{Leptonic-Hadronic Origin Model}\label{sec:hadron}
For the hadronic process, it is thought that the neutron star atmosphere consists mainly of irons \citep{1997ApJ...484..323V}. The Fe nuclei of the neutron star surface can escape from the polar cap surface of the pulsar and move along magnetic field lines to enter the pulsar magnetosphere where those heavy nuclei can be accelerated by the potential \citep{1990JPhG...16.1115C,1997PhRvL..79.2616B}. After the acceleration of the pulsar magnetosphere, the partial particles will escape from the pulsar magnetosphere along the open magnetic field lines and be injected into the PWN. During the propagation process, the Fe nuclei and the other heavy nuclei will interact with the background photons and suffer disintegration in the collisions with soft photons produced inside the pulsar outer gap. The injected rate of  Fe nuclei was given as \citep{1997PhRvL..79.2616B}

\begin{equation}
\label{equation3}
 \dot{N}_{Fe}=\xi L_{\rm Crab}(B,P)/Z\Phi(B,P)
\end{equation}
where $\xi$ is the fraction of the total power taken by relativistic nuclei accelerated in the outer gap; $Z$ is the atomic number of Fe. And the potential difference across the outer gap can be described simply as $\Phi(B,P)\approx5\times10^{16}(B/4\times10^{12}G)(P/s)^{4/3} \rm V$, $B$ is the surface magnetic field of the pulsar.

The protons emitted directly in the disintegration process and those protons from neutron decay in the evolution process are captured by the nebula and accumulated in the PWNe. Therefore, the proton spectrum was described as
\begin{equation}
\label{equation4}
 \frac{dN_{p}}{dE}=\bigg[\frac{dN_{p}(\gamma_{p},t_{CN})}{dE}\bigg]_{1}+\bigg[\frac{dN_{p}(\gamma_{p},t_{CN})}{dE}\bigg]_{2}
\end{equation}
Here, the proton spectrum from the neutron decaying is \citep{1997PhRvL..79.2616B}
\begin{equation}\label{eq:mti}
  \begin{split}
    \bigg[\frac{dN_{p}(\gamma_{p},t_{CN})}{dE}\bigg]_{1}=\gamma_{p}^{-1}\int^{t_{CN}}_{0}dt\dot{N}_{Fe}(t)e^{-\tau_{nH}(t)}\\
    \times\bigg[N_{n}(\gamma_{p}(t),t)\gamma_{p}(t)[1-exp(-vt/c\gamma_{p}(t)\tau_{n})]+\\
    \int^{t_{CN}}_{t}dt'N_{n}(\gamma_{p}(t'),t)\gamma_{p}(t')\frac{vexp(-vt'/c\gamma_{p}(t')\tau_{n})}{c\gamma_{p}(t')\tau_{n}}\bigg]
  \end{split}
\end{equation}
and the direct proton spectrum derived from the photodisintegration of the Fe nuclei and the other heavy nuclei is \citep{2009A&A...496..751Y}
\begin{equation}\label{eq:mti2}
  \begin{split}
    \bigg[\frac{dN_{p}(\gamma_{p},t_{CN})}{dE}\bigg]_{2}=\frac{\gamma_{p}(t)}{\gamma_{p}}\dot{N}_{Fe}(t)\frac{dt}{dP}\frac{dP}{dE_{p}(t)}\\
    e^{-\tau_{nH}(t)}\int^{\gamma_{p}(t)}_{0}d\gamma'N_{p}(\gamma')
  \end{split}
\end{equation}
where number of the $i$ nucleons at energy $E_{i}$ ($i=n,p$) per unit energy per one nucleus was approximated $N_{n,p}(E_{n,p})\approx\frac{Al_{gap}R_{A,i}}{Z\Phi(B,P)c}$ , the $l_{\rm gap}$ is the dimension of the outer gap, $R_{A,i}$ is the photodisintegration rate of nuclei. Its calculative detail in the PWNe was described in \cite{1997PhRvL..79.2616B}. Here $A$ is atomic number of nucleus and $\gamma_{p}(t)$ presents the Lorentz factor at time $t$. The $\tau_{nH}(t)$ and $\tau_{n}$ present the optical depth and neutron decaying time, respectively. In the calculated process, we adopt the $\xi=0.85$ fitted in \cite{2003A&A...407....1B}.

The protons provided in the above are accelerated by the pulsar magnetosphere. However, the emission with energy extended to the $\rm PeV$ has been observed through the LHAASO \citep{2021Natur.594...33C} and this indicates that the primary particles need to reach the energy of exceeding $\rm PeV$. But the iron nuclei accelerated in the inner magnetosphere or the Crab pulsar wind zone may mainly contribute in the $\rm TeV$ gamma-ray range \citep{1990JPhG...16.1115C,2003A&A...407....1B}. In this paper, we assume an extra component of the proton spectrum as the exponential cutoff power-law distribution
\begin{equation}
\label{equation7}
\frac{dN}{dE}=A_{p}E^{-\alpha_{p}}_{p}exp(-\frac{E}{E_{p,c}})
\end{equation}
Here, $A_{p}$ is the normalization factor. We constrain the parameter $A_{p}$ with the observed data in the sub-$\rm PeV$ to $\rm PeV$ energy range. The $E_{p,c}$ is cutoff energy of protons.
According to the discussion in some literature, the protons can be accelerated up to an energy above 30 PeV \citep{2000ApJ...533L.123B,2020MNRAS.497.3477Z,2011MNRAS.410..381B}.
Some acceleration sites in the Crab nebula may provide a source of ultrarelativistic light protons with an injection  spectrum, steepening to $E^{-2}$ at higher energies \citep{2003ApJ...589..871A}. Here, we fix the index $\alpha_{p}=2.0$  and set $E_{p,c}= 30~\rm PeV$ due to constraint with few $\rm PeV$ data. We adopt a mean gas density, which is estimated as $n_{H}\sim10~\rm cm^{-3}$ for the Crab nebula with the radius $R_{\rm pwn}=1.8~\rm pc$ \citep{2020MNRAS.497.3477Z,2021Sci...373..425L}.

\begin{figure*}[t]
\centering
\begin{minipage}{\textwidth}
\includegraphics[width=0.5\textwidth]{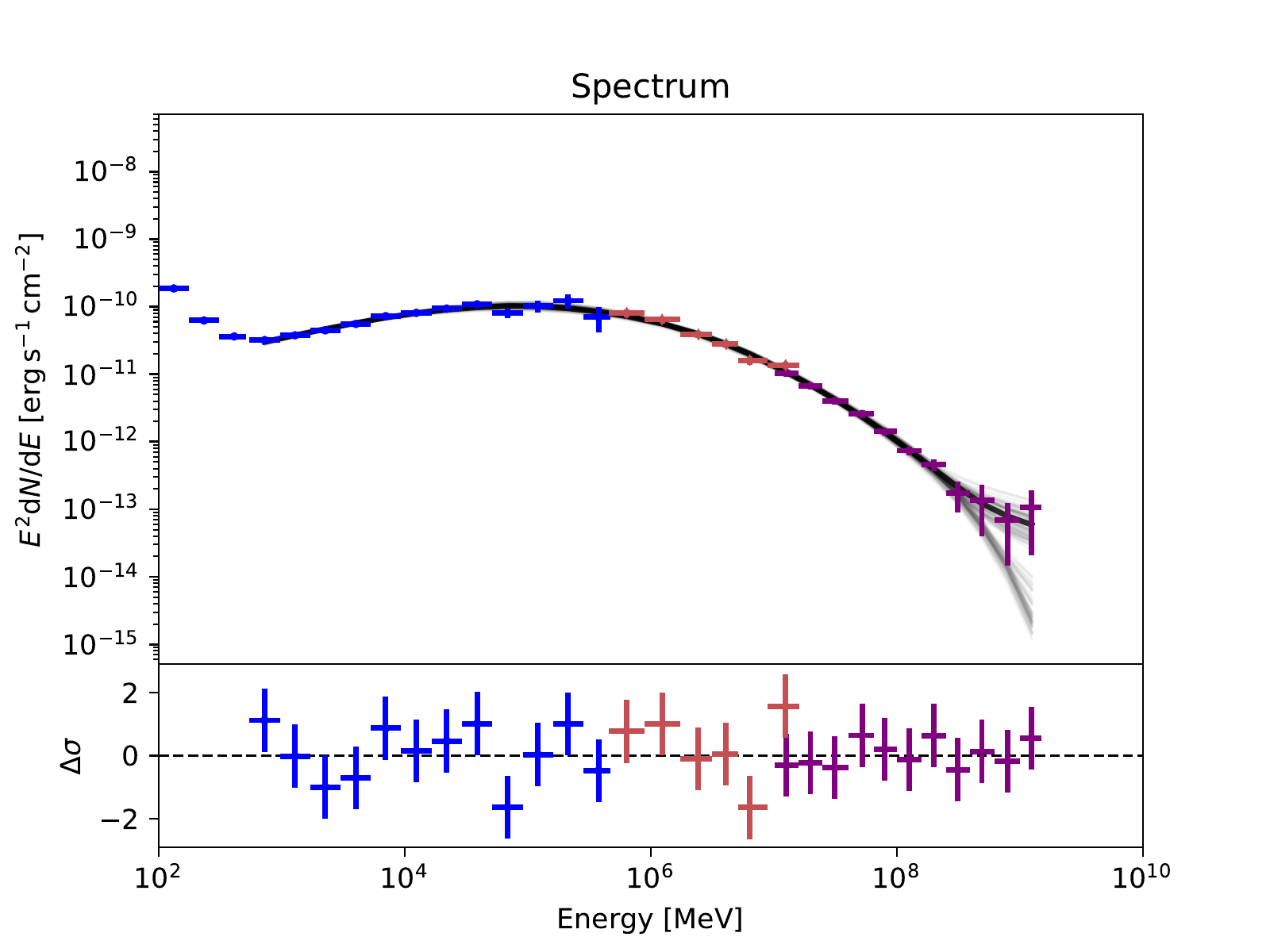}
\includegraphics[width=0.5\textwidth]{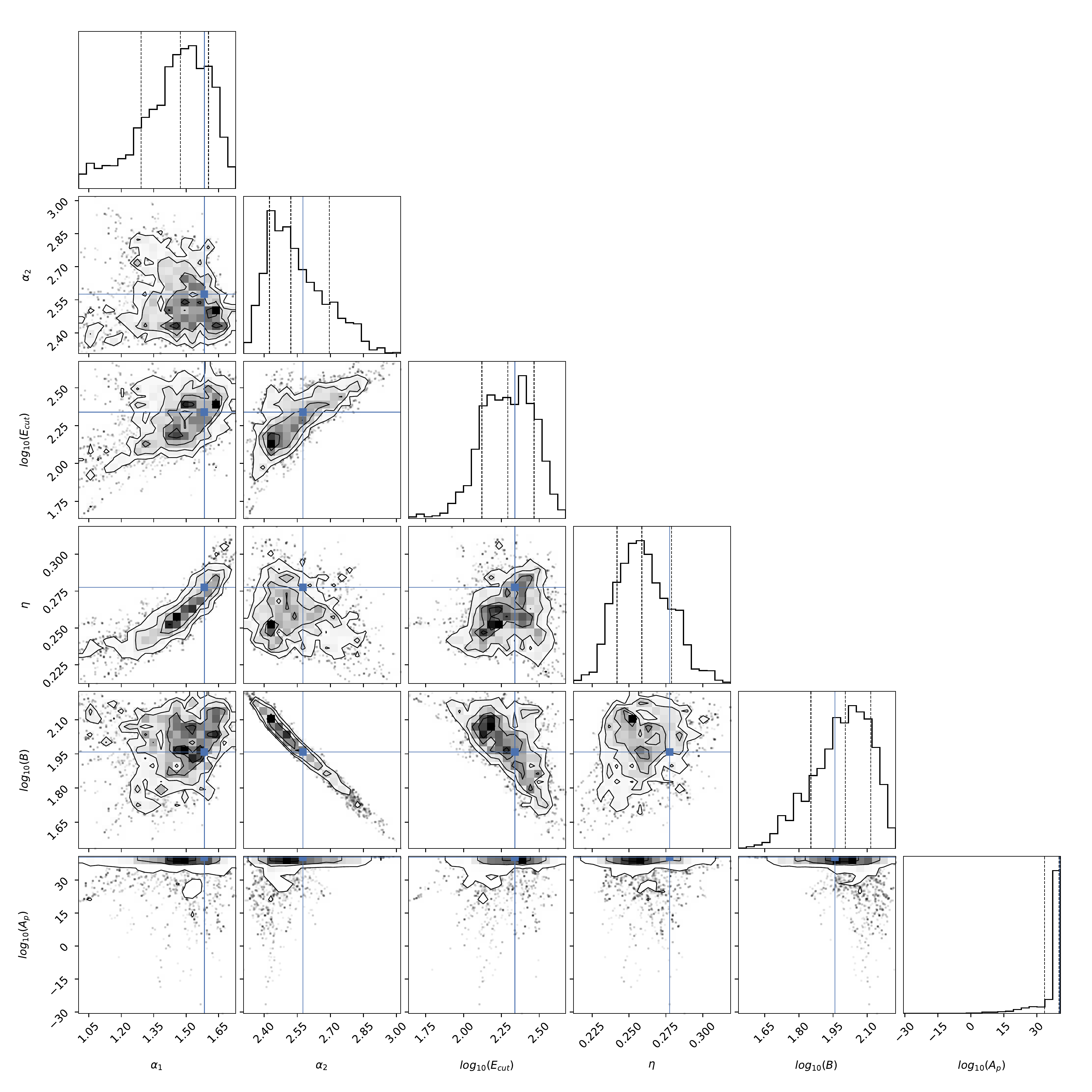}
\end{minipage}
\caption{The same as Fig.\ref{Figure1} but the hadronic and bremsstrahlung components have been considered. This hybrid model adds a free parameters ($A_{p}$) and ignores the parameter $\epsilon_{0}$ due to setting the exponential cutoff energy of electrons to $E_{0}=450~\rm TeV$.}
\label{Figure3}
\end{figure*}

\begin{figure*}[t]
\centering
\begin{minipage}{\textwidth}
\includegraphics[width=0.5\textwidth]{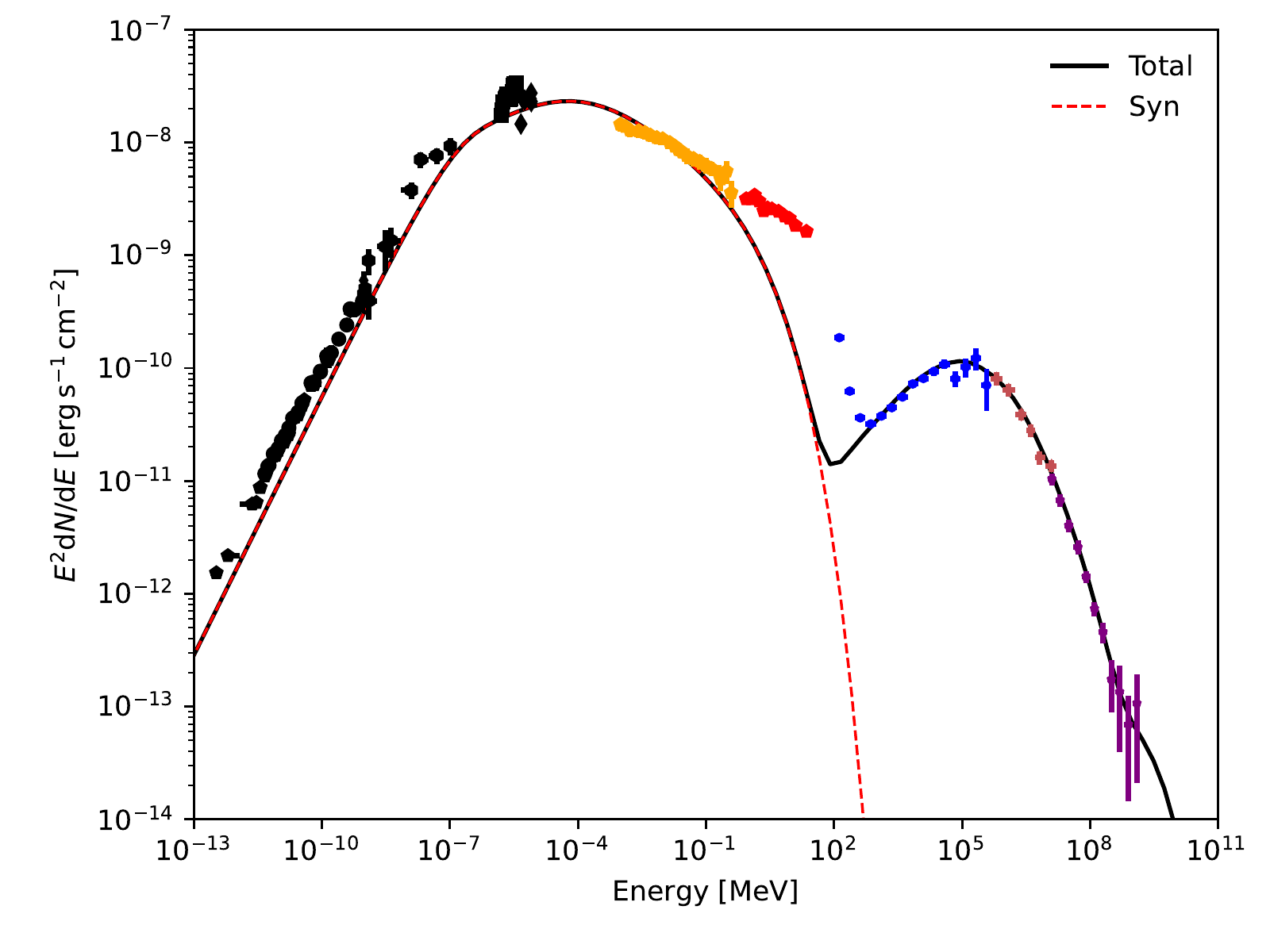}
\includegraphics[width=0.5\textwidth]{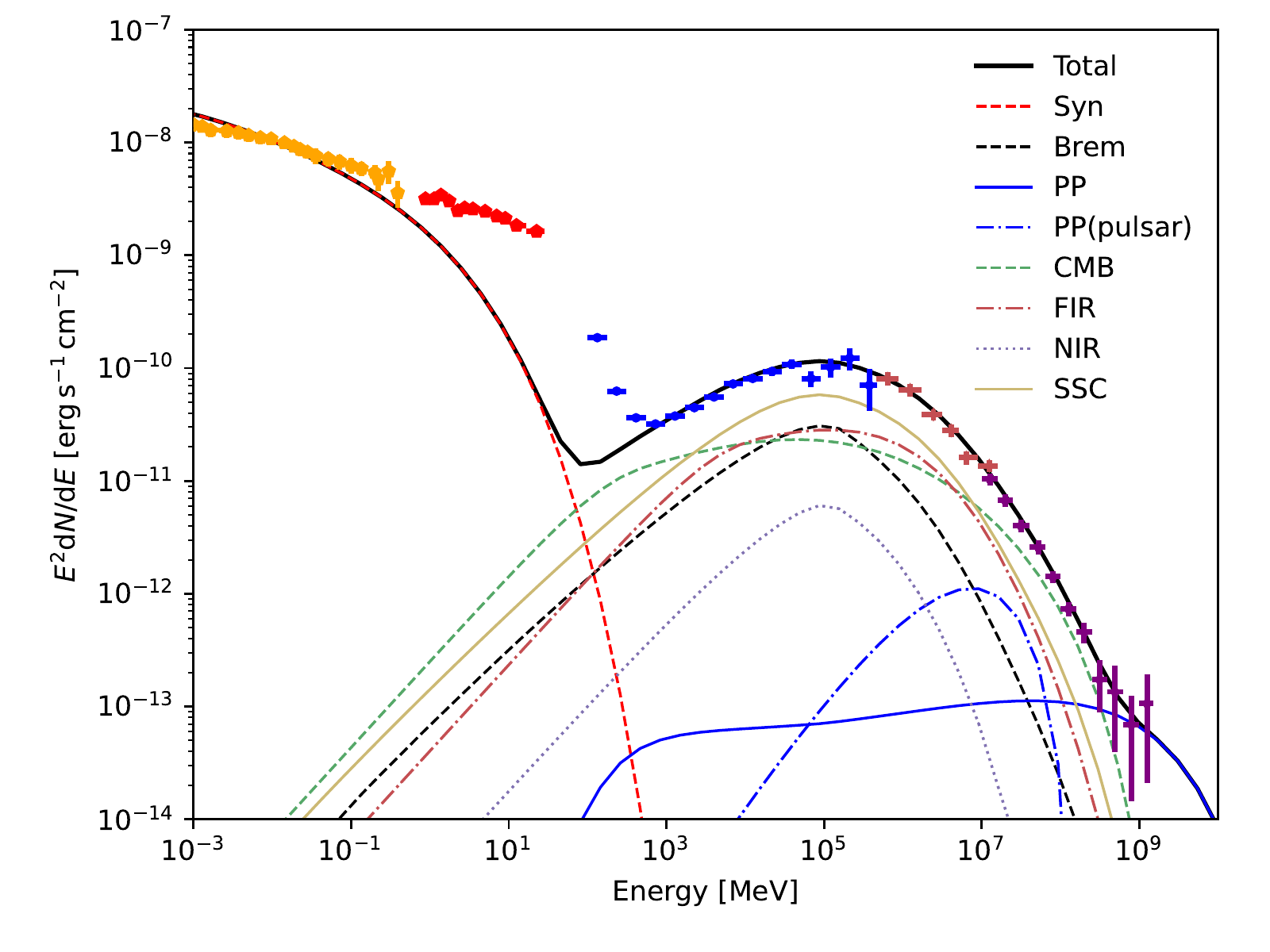}
\end{minipage}
\caption{The same as Fig.\ref{Figure2} but the hadronic and bremsstrahlung components have been considered. Here, the blue dashed-doted line presents the component of protons that originates from the photodisintegration of Fe in the pulsar magnetosphere; the blue solid line is an extra component from the protons. The black dashed line shows the bremsstrahlung of electrons.}
\label{Figure4}
\end{figure*}

In order to explain the multiwavelength observational results from the PWNe, the contribution provided by the leptons is essential. The electrons generate photons through synchrotron, bremsstrahlung, and IC scattering mechanisms; and protons produce the gamma rays via the decay of $\pi^{0}$ mesons in proton$-$proton interactions. Here, based on the distribution of electrons and protons, we calculated the multiband nonthermal emission. The gamma-ray radiation from proton$-$proton interaction is calculated using the expressions of \cite{2006PhRvD..74c4018K} and the bremsstrahlung is calculated via the formalism of \cite{1970RvMP...42..237B}.

However, the model assumes that the emission is produced in an idealized one-zone bulk in Section \ref{sec:lepton}. This estimate is based on the assumption that the magnetic field has a uniform distribution or is slowly varying. Furthermore, the numerical results of PWNe have shown clearly that the distribution of the magnetic field and the maximum energy of the emitting particles can change quite dramatically within the nebula \citep{2008A&A...485..337V}. The $\rm MeV$ synchrotron radiation and sub-$\rm PeV$ IC emission are mainly produced by the electron population with energy exceeding $\thicksim450~\rm TeV$ \citep[see, e.g.,][]{2020MNRAS.491.3217K,2021Sci...373..425L}. On the other hand, the structure that the gamma-ray spectrum in the $1-100~\rm MeV$ band is not smooth can be reproduced by the electron population with an exponential cutoff $E_{\rm cut}\thicksim450~\rm TeV$ and a hard energy distribution peaking at higher energies \citep{2020MNRAS.491.3217K}.
Therefore, we set the electron population to have an exponential cutoff energy $E_{0}=450~\rm TeV$, assuming the $\rm MeV$ synchrotron radiation and $\rm PeV$ IC emission have different origins. Then we use the new component from the proton to provide the extra $\rm PeV$ emission. We mainly focus on the high-energy emission, namely the $\rm GeV$ to $\rm PeV$ emission. Here, the data with energy below $\rm GeV$, which may have a more complicated origin, is ignored during the fitting process.

It is similar to the treatment method of the one-zone leptonic model in Section \ref{sec:lepton}. Together with the leptonic scenario introduced in Section \ref{sec:lepton}, we refit the observed data ranging from $\rm GeV$ to $\rm PeV$ gammarays for the Crab nebula with the MCMC approach.
As results of the analysis, we show statistics of the parameter distribution, which includes a medium with uncertainties based on the 16th and 84th percentiles in Fig.\ref{Figure7} (see the Appendix.\ref{appendix A}).
we also show the corner plot, which plots the distribution for all parameters against each other and the spectral energy distribution with the best-fitting parameters in Fig.\ref{Figure3} and Fig.\ref{Figure4}.
The results present that this electron population underestimates the fluxes in the radio to ultraviolet (UV) bands. This is in favor of the scenario that the radio$-$UV emission has a more complicated mechanism, which is consistent with the previous investigation \citep{2019MNRAS.489.2403L}. But the data and model have a smaller deviation in the energy range above $\rm GeV$, compared with the one-zone leptonic situation in Section \ref{sec:lepton}. And the data can be described better.

As a result of the calculation, the best-fitting parameters of the leptonic$-$hadronic model are listed in Table.\ref{tab:2}. The fitting values of some parameters that have been fitted with the MCMC in Section \ref{sec:lepton} have a significant change. Finally, we obtain the total energy in protons inside the Crab nebula $W_{p}=3.64^{+4.08}_{-3.6}\times10^{47} \rm erg$.

\begin{table*}
\tablenum{2}
\begin{center}
\caption{The best-fitting parameters for the leptonic$-$hadronic model.}
\scriptsize
\renewcommand\tabcolsep{2.0pt}
\renewcommand\arraystretch{1.1}
\begin{tabular}{@{}ccccccc@{}}
\label{tab:2}\\
\hline\hline
 Model& $ \alpha_{1} $ & $\alpha_{2}$ & $ \eta $ & $B_{\rm PWN}(uG)$ & $E_{\rm cut}(\rm TeV)$ & $A_{p}(eV^{-1})$\\
\hline
leptonic$-$hadronic& $1.49^{+0.13}_{-0.18}$ & $2.53^{+0.13}_{-0.09}$& $0.259^{+0.02}_{-0.015}$ &$102.33^{+23.56}_{-6.83}$ & $0.204^{+0.071}_{-0.075}$& $1.58^{+1.57}_{-1.56} \times10^{40}$\\
\hline
\hline
\end{tabular}
\end{center}
\end{table*}

\begin{figure}[t]
\centering
\begin{minipage}{\textwidth}
\includegraphics[width=0.45\textwidth]{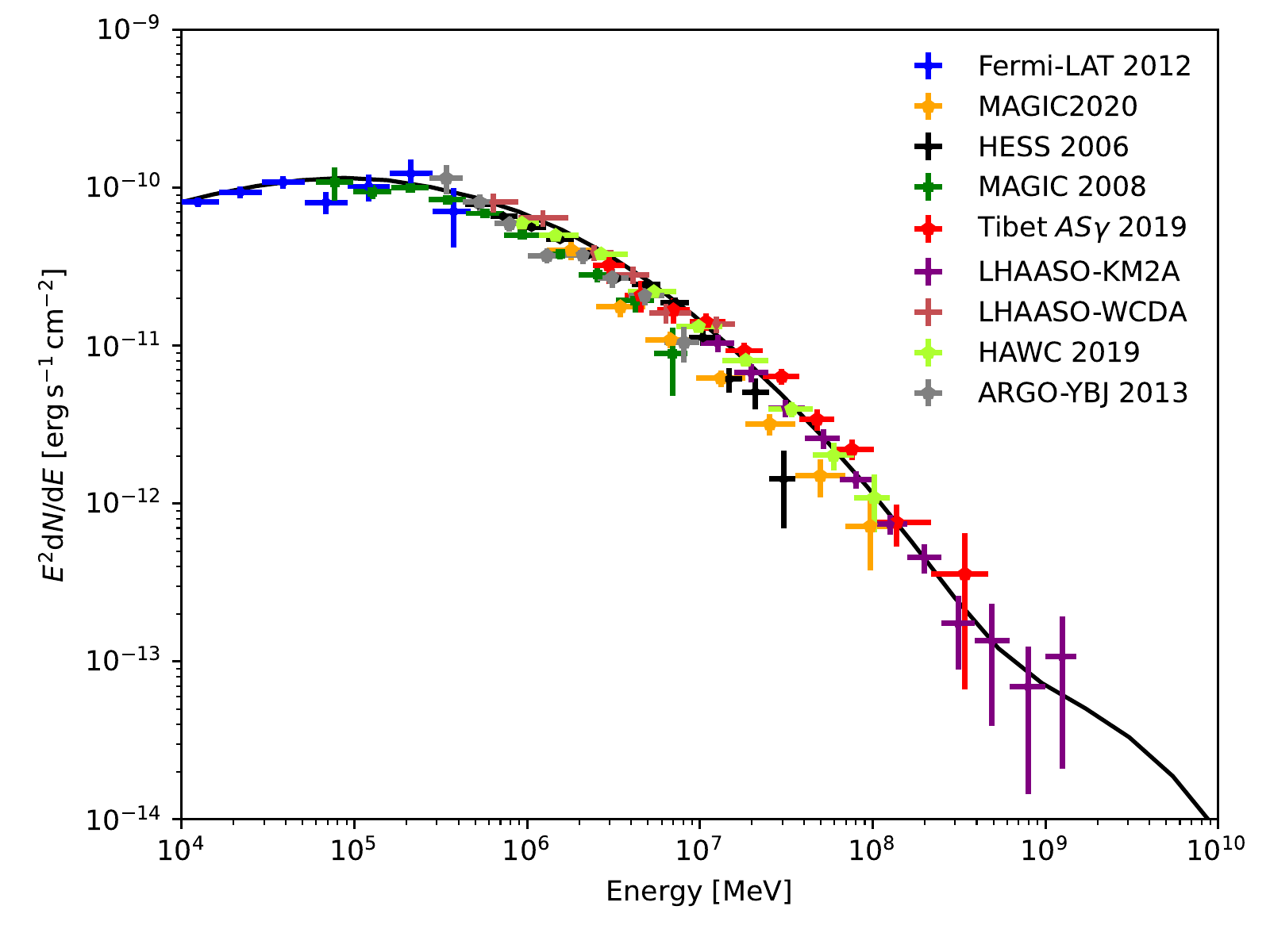}
\end{minipage}
\caption{Gamma-ray spectrum of the Crab Nebula. The gamma-ray data are taken from the High-Energy Stereoscopic system (H.E.S.S;\cite{2006A&A...457..899A}), the Major Atmospheric Gamma Imaging Cerenkov (MAGIC) telescopes \citep{2020A&A...635A.158M}, the High Altitude Water Cherenkov (HAWC) Gamma-ray Observatory \citep{2019ApJ...881..134A}, the Fermi Large Area Telescope \citep{2020ApJ...897...33A}, Astrophysical Radiation by Ground-based Observation at Yang Ba Jing \citep{2015ApJ...798..119B}, Tibet AS$\gamma$ \citep{2019PhRvL.123e1101A}, and LHAASO \citep{2021Sci...373..425L}. The black solid line is the same as the one in the left panel of Fig.\ref{Figure3}.}
\label{figure5}
\end{figure}

\section{Conclusion and Discussion} \label{se:conclusion and discussion}
The nonthermal emission from the Crab nebula, with the energy ranging from radio to high-energy gamma ray bands, is emitted by the injected electrons through synchrotron emission and IC scattering mechanism, or is also produced via the decay of $\pi^{0}$ mesons in the proton$-$proton interaction process. Based on the previous investigation, the observational data ranging from radio to high-energy gamma rays reported by the different observatories is generally interpreted as a strong argument in favor of a leptonic origin \citep[see, e.g.,][]{2012MNRAS.427..415M, 2013ApJ...763L...4T, 2015MNRAS.451.3145Z}. On the other hand, the hadronic origin is also naturally used to account for the high-energy gamma-ray emission \citep[see, e.g.,][]{1997PhRvL..79.2616B,2003A&A...407....1B}. Therefore, it is argued that it is difficult to distinguish the leptonic and hadronic origin of the emission with energy exceeding $\sim10~\rm TeV$ for the Crab nebula \citep[see, e.g.,][]{2003A&A...407....1B, 2009A&A...496..751Y, 2020MNRAS.491.3217K, 2020MNRAS.497.3477Z}.

In order to calculate the multiwave band nonthermal radiation from the PWNe, we need to determine the basic parameters of pulsar and nebula. For the Crab nebula, the break index $n=2.5$, initial period $P_{0}=19~\rm ms$ and moment of inertia $I=1.0\times10^{45}~\rm g~cm^{2}$ have been known \citep{2003A&A...407....1B, 2007Ap&SS.309..179B,2008ApJ...676.1210Z}. In this paper, we model the multiwavelength nonthermal radiation from the Crab nebula with the leptonic origin model and leptonic$-$hadronic hybrid model, respectively. In addition, we use the method of Markov Chain Monte Carlo sampling to obtain the value of the maximum likelihood parameters and their uncertainties.

We find out that the leptonic model with the one-zone fails to describe the broadband observed data in detail, and the hadronic component may be crucial for gamma-ray emission from the Crab nebula. This is because, in some energy range, the emission may have a different origin. For instance, the $\thicksim100~\rm keV - MeV$ emission may arise from a superposition of two components \citep{2019MNRAS.489.2403L}. With the report of gamma-ray flares ($100~\rm MeV-10~GeV$) from the Crab nebula \citep{2013ApJ...775L..37M,2020ApJ...897...33A}, it is thought that the $\rm MeV$ radiation might be produced in a special region with special physical conditions (e.g. the particles are accelerated by the magnetic reconnection or a small-scale magnetic turbulence is present) which are different from the typical conditions expected in the Crab nebula \citep{2012ApJ...754L..33C,2013ApJ...774...61K,2020ApJ...896..147L}. The radio to UV emission is the most complicated and its resulting broadband SEDs include synchrotron emission of the accumulated long-living leptons, thermal emission from the dust in the nebula, and optical line emission from the filaments \citep{2010A&A...523A...2M}. Therefore, when we fit the global data within the one-zone leptonic model, the constraint to the model is mathematically suppressed and some segments of data show a strong signal of deviation from the model.
For the VHE gamma-ray emission, although we cannot distinguish whether it is dominated by the leptonic or hadronic origin, it is responsible to constrain some parameters together with the contribution provided by the hadronic interaction process.

On the other hand, the gamma-ray emission beyond $\rm PeV$ could need an extra radiative component.
This is because, although the spectrum simulated via the one-zone leptonic model seems to explain the observational $\rm PeV$ data (see Fig.\ref{Figure1}), the observational data of around $1.1~\rm PeV$ with the highest energy reported by the LHAASO \citep{2021Sci...373..425L} have a higher flux than that of the prediction. Importantly, the acceleration rate of emitting electrons at such high energy is higher than $21\%$. This could be a  huge challenge to the ideal MHD or even classical electrodynamics. While the leptonic$-$hadronic scenario perhaps is a reasonable result. It has a smaller deviation (see Fig.\ref{Figure4}). In addition, the $\rm PeV$ emission was also explained as an extra exponent from leptons, which originates from regions with $B<100~\rm uG$ \citep{2021Sci...373..425L}. But it remains to be explored whether the $\rm PeV$ photons are dominated by the electrons and we leave it to future studies.

We show the previous measurements in Fig.\ref{figure5}. They are almost consistent with the model and the observation of LHAASO. Yet there seems to be a deviation from the model for the measurement of different experiments. The data of HAWC and Tibet AS$\gamma$ have a smaller deviation that is consistent with the WCDA and KM2A data. While the data from H.E.S.S, MAGIC, and ARGO-YBJ are lower than the prediction of the model. This could be due to the experiments with  different systematic uncertainty.

In the fitting process of the leptonic$-$hadronic hybrid model, base on the constraint of the current data ranging from $\rm GeV$ to ultra-high-energy gamma-rays which have been reported, we calculate the energy of exceeding $\rm PeV$ in protons as $W_{p,>PeV}\sim5.95^{+7.16}_{-5.89}\times10^{46} \rm erg$  inside the Crab nebula. We find that the contribution of hadronic interaction is hardly constrained.
It should be remarked that the parameter ($A_{p}$) is not stabilized very well in the fitting process (see Fig.\ref{Figure7}) due to few observational data with energy extended to the $\rm PeV$. It is mainly constrained by the gamma-ray data of extending $\rm PeV$. Therefore, it needs further observation to provide constraints or evidence for the contribution of hadronic interaction in the future.

\section*{Acknowledgements}
We thank the referee for providing some suggestions. This work is partially supported by the National Key Research and Development Program 2018YFA0404204, the National Natural Science Foundation of China (NSFC U1931113, U1738211), and the Foundations of Yunnan Province (2018FY001(-003)).

\appendix
\section{The Stability of Model Parameters around the Best-fitting Values}\label{appendix A}
\begin{figure}[h]
\centering
\begin{minipage}{\textwidth}
\includegraphics[width=0.49\textwidth]{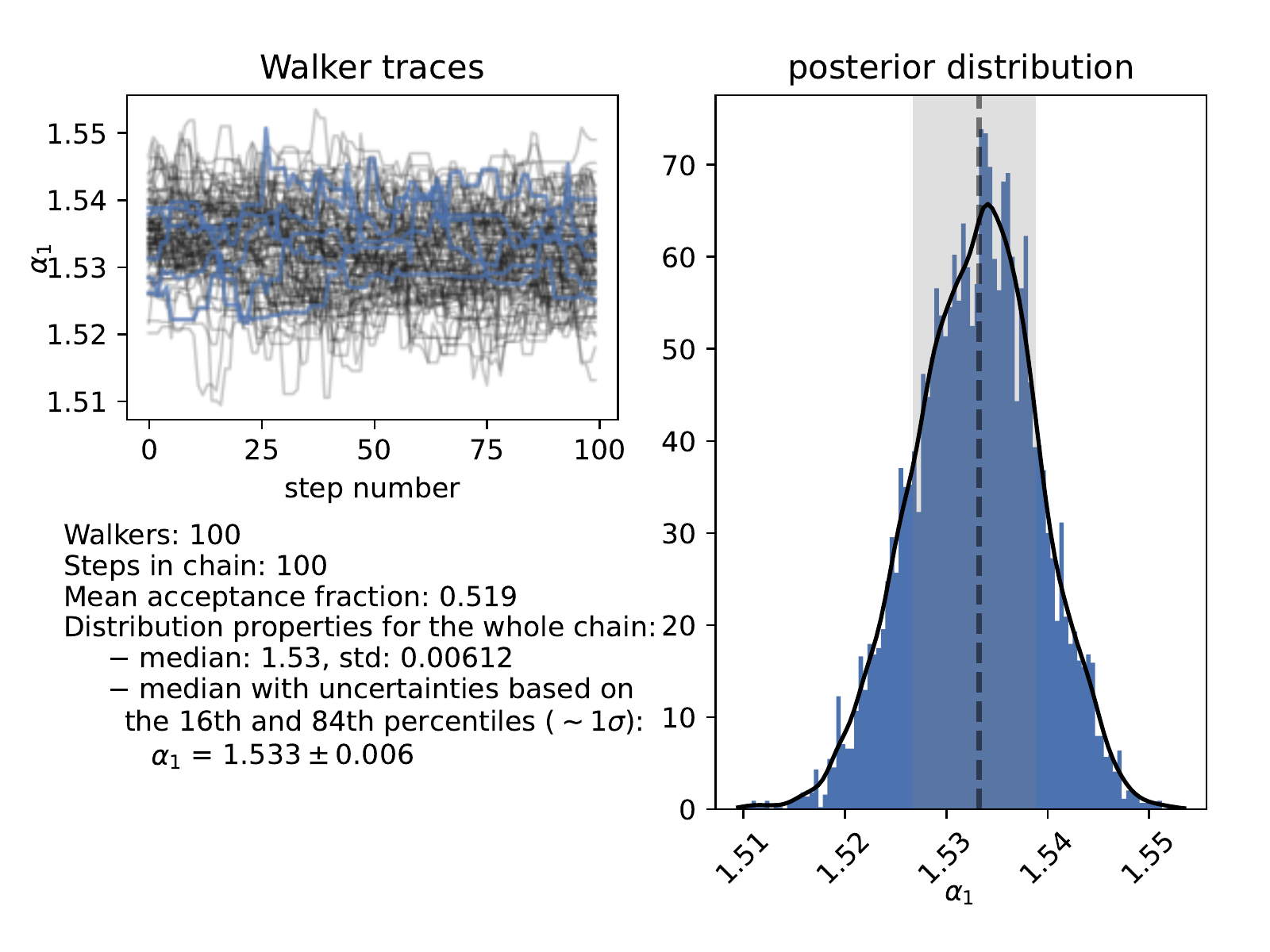}
\includegraphics[width=0.49\textwidth]{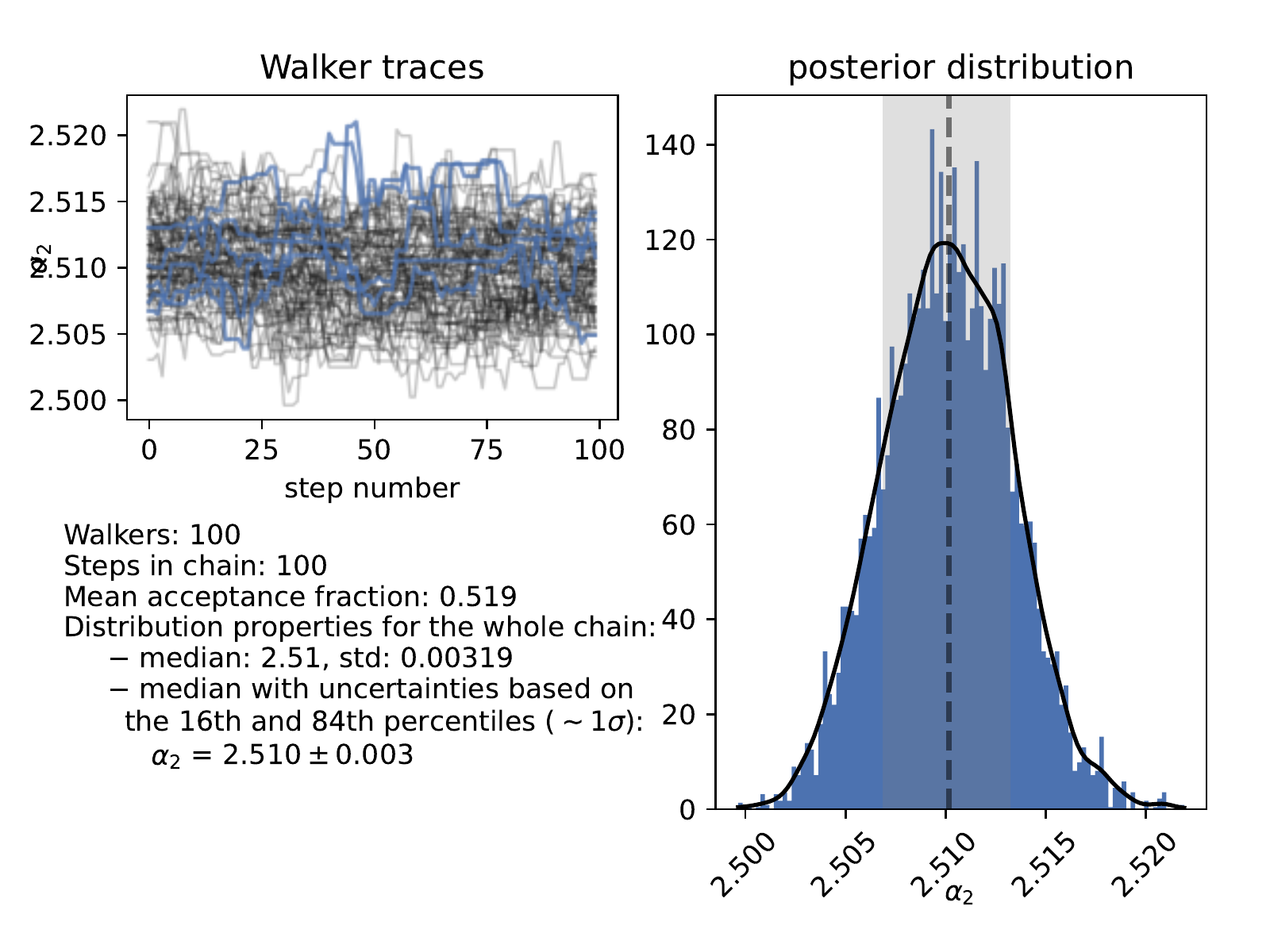}
\includegraphics[width=0.49\textwidth]{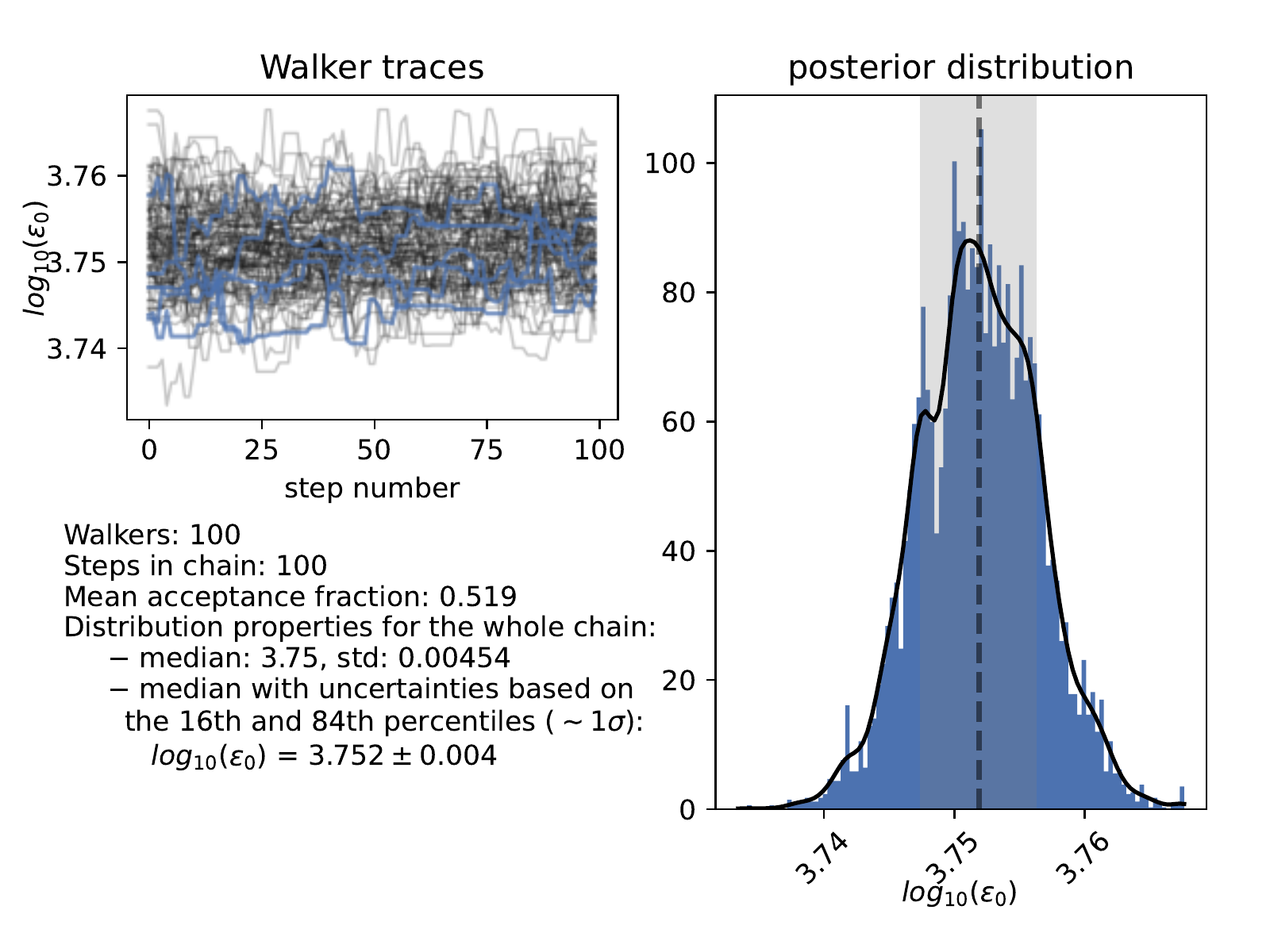}
\includegraphics[width=0.49\textwidth]{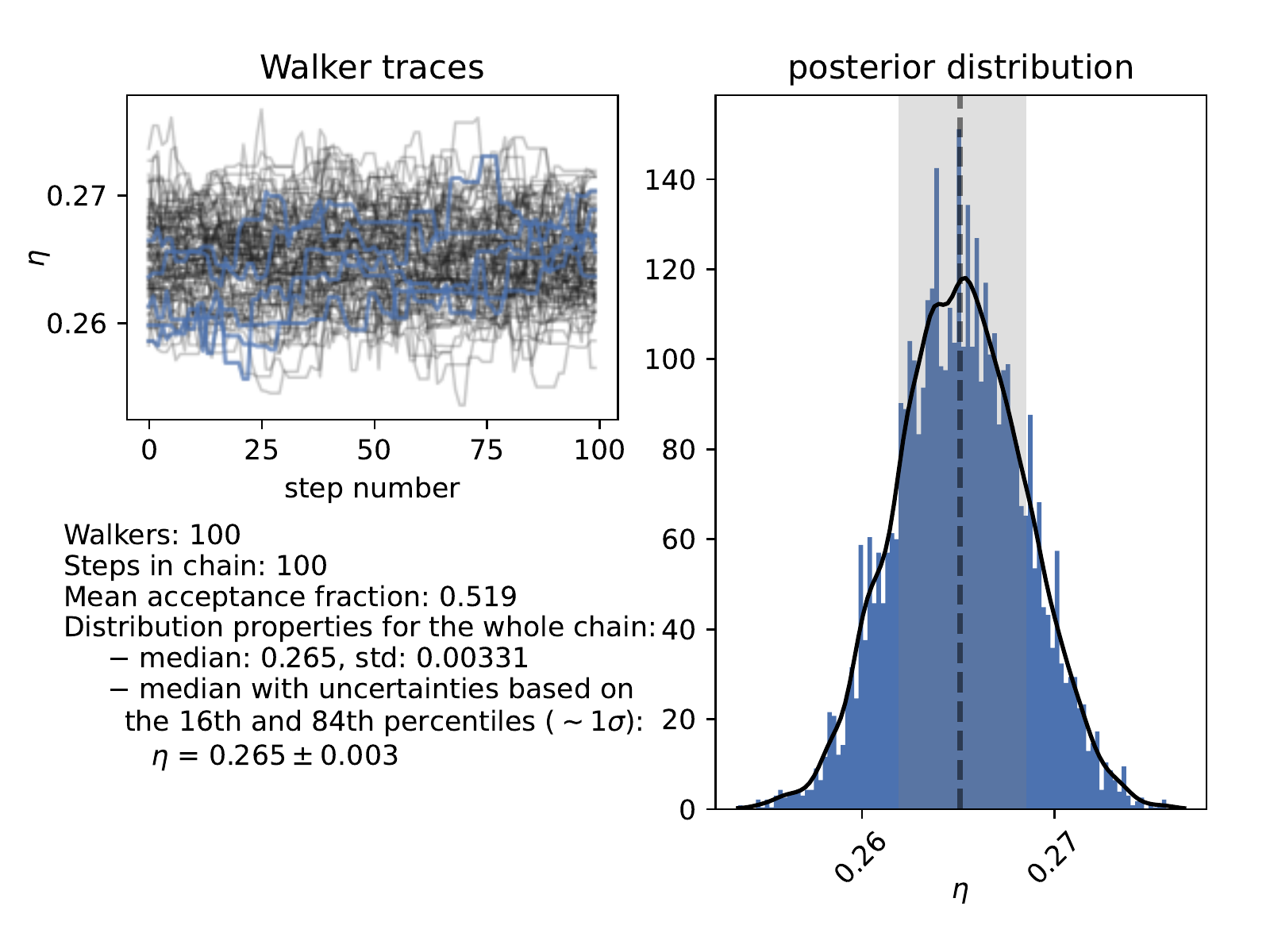}
\end{minipage}
\caption{The diagnostic figure in the sampling of the parameters in the leptonic model. The top-left panel shows the traces for the 100 workers in gray and three of them are highlighted in blue in each panel, which can be used to estimate whether the sampling has stabilized around the maximum likelihood parameters. The right panel shows the posterior distribution of the parameters in the individual plot. (Figure 6 continued on next page)}
\end{figure}

\setcounter{figure}{5}
\begin{figure}[h]
\centering
\begin{minipage}{\textwidth}
\includegraphics[width=0.49\textwidth]{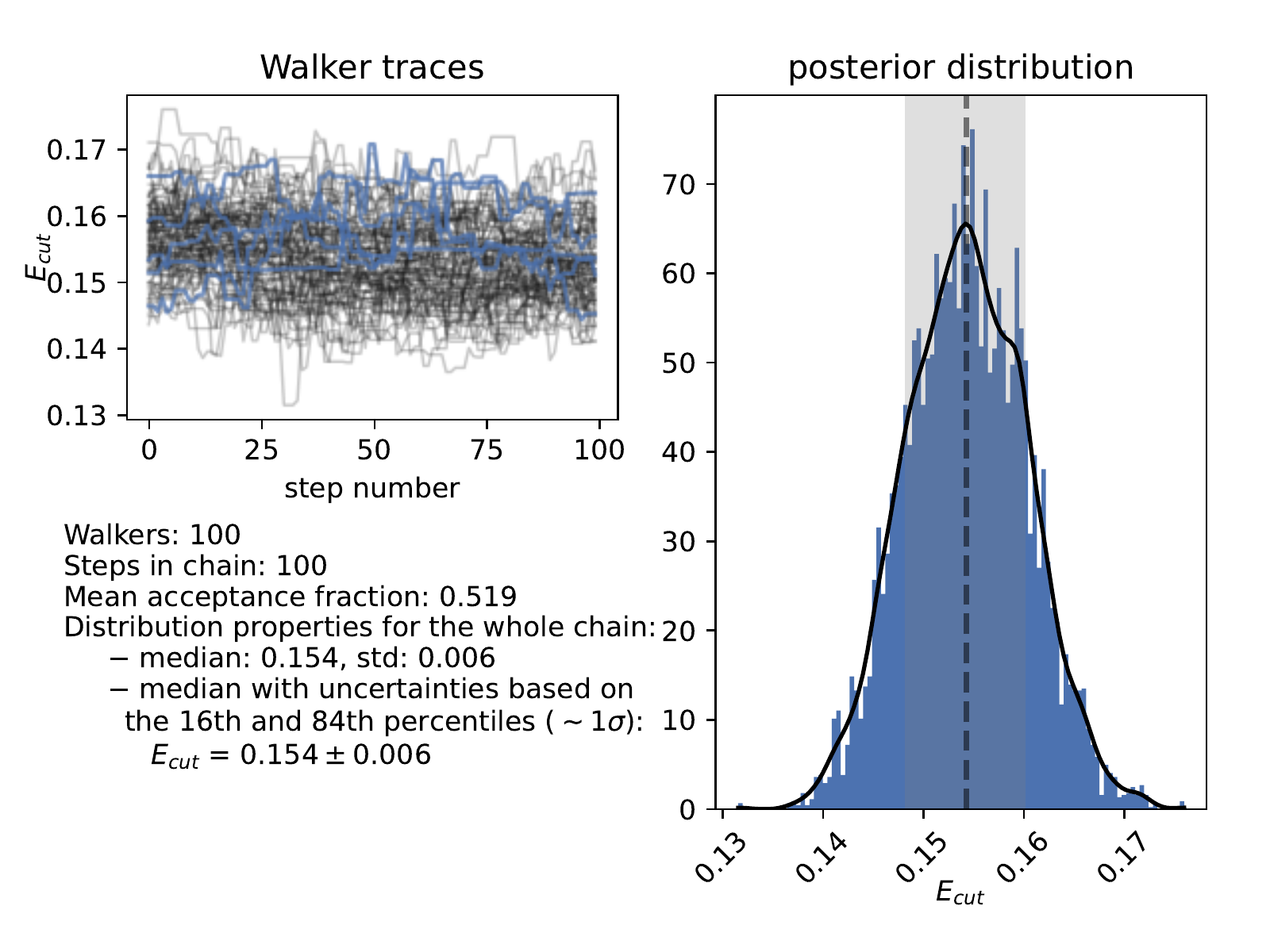}
\includegraphics[width=0.49\textwidth]{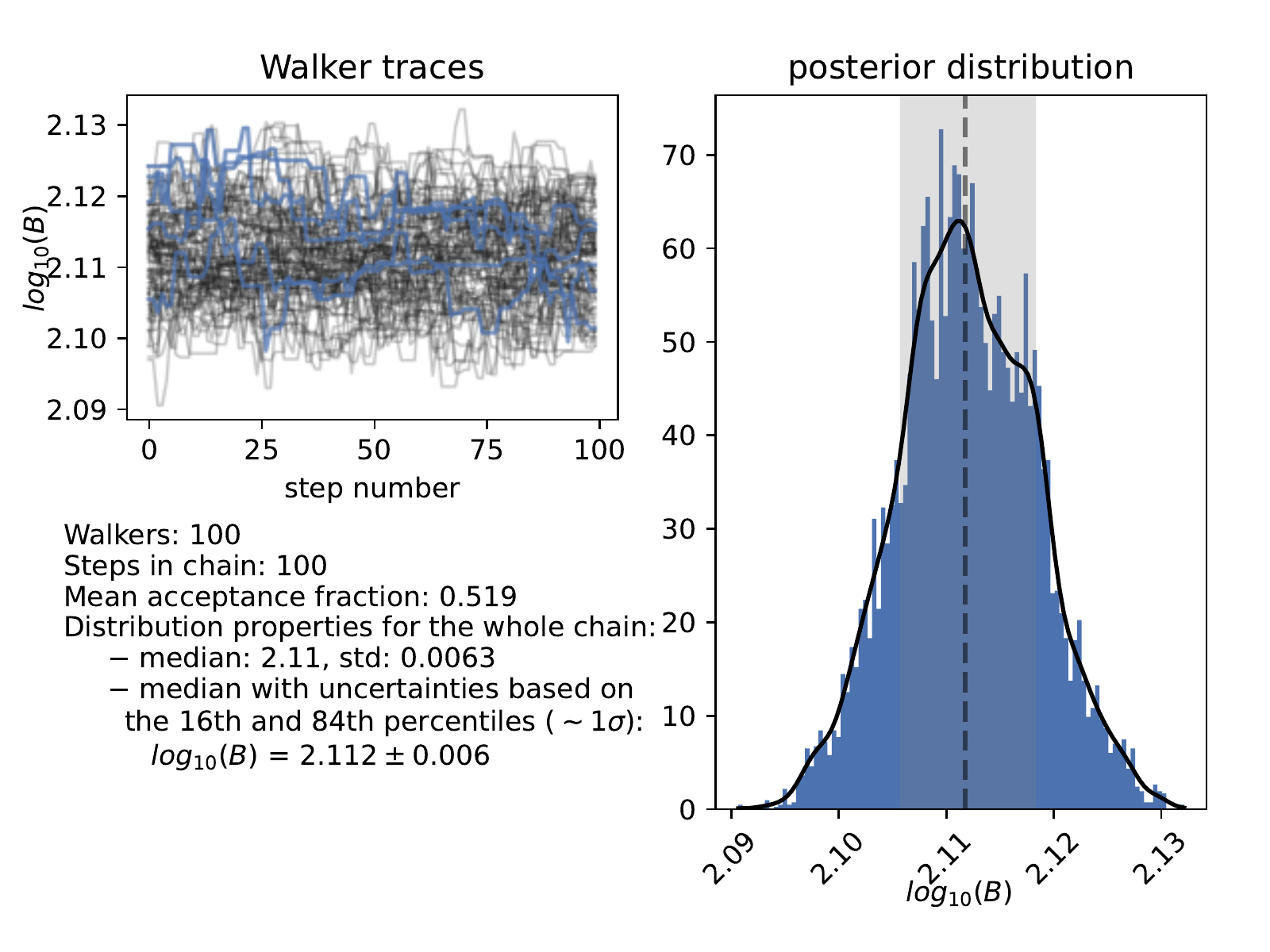}
\end{minipage}
\caption{(continued)}
\label{Figure6}
\end{figure}

\begin{figure}[h]
\centering
\begin{minipage}{\textwidth}
\includegraphics[width=0.49\textwidth]{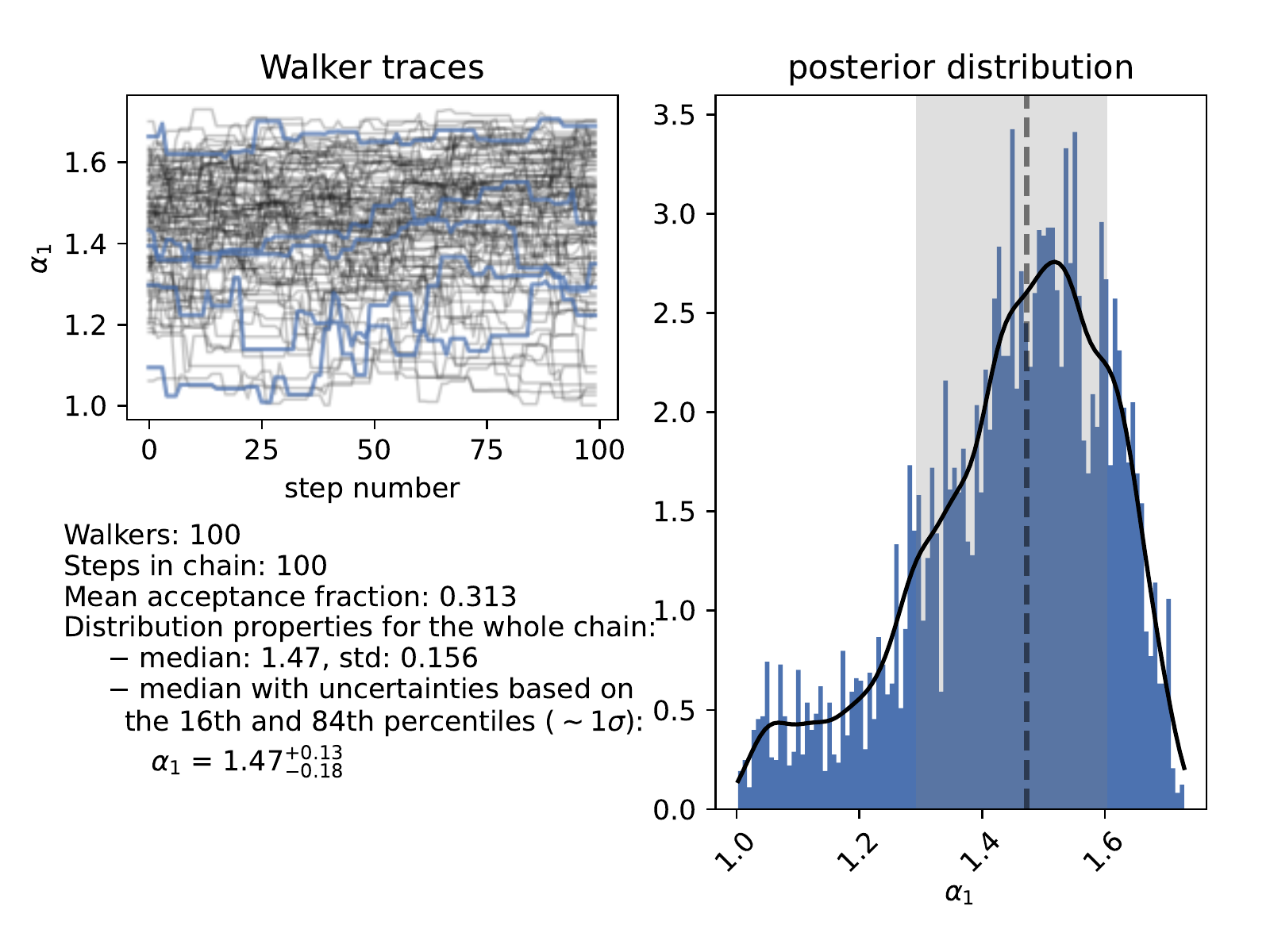}
\includegraphics[width=0.49\textwidth]{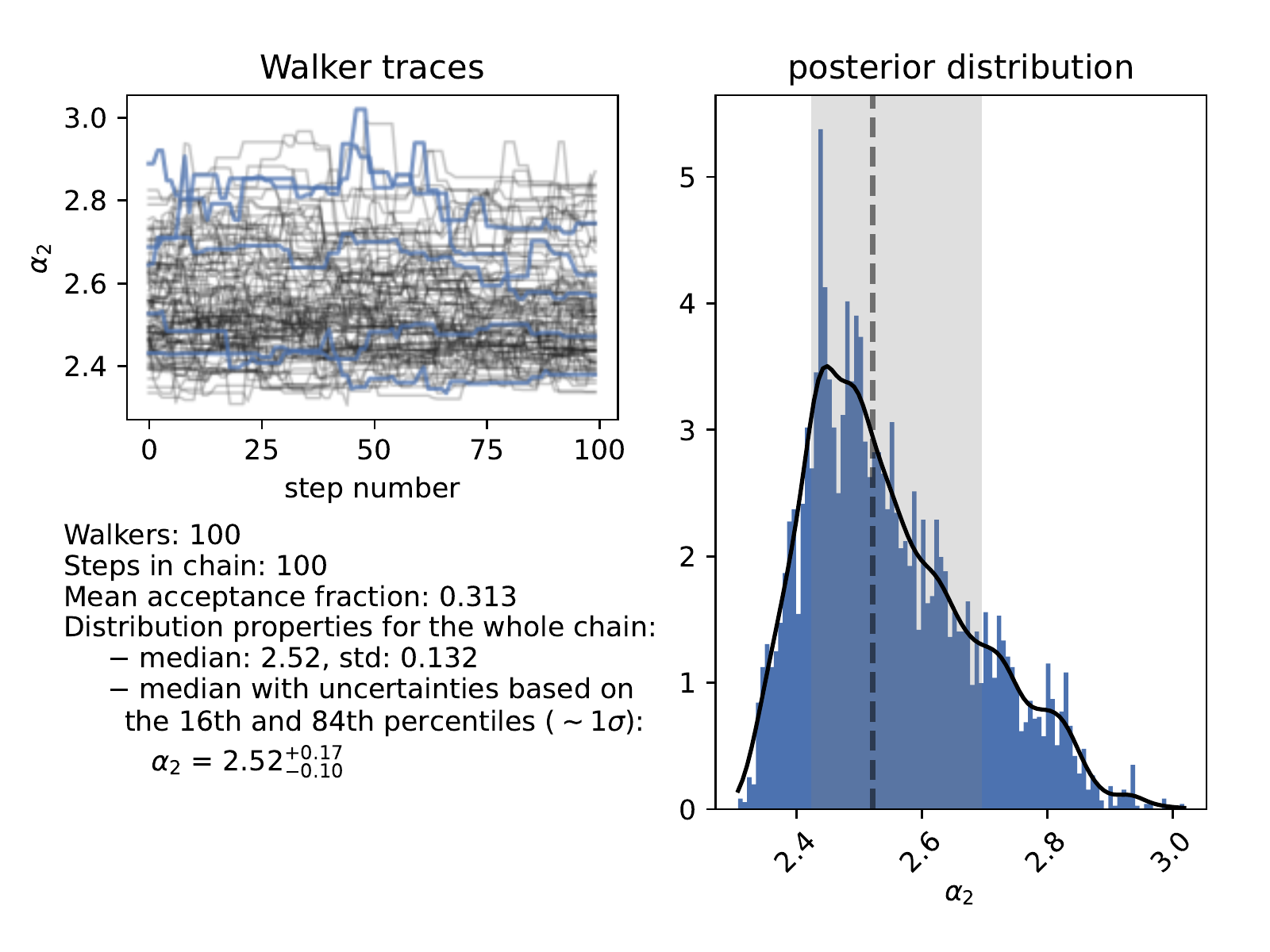}
\includegraphics[width=0.49\textwidth]{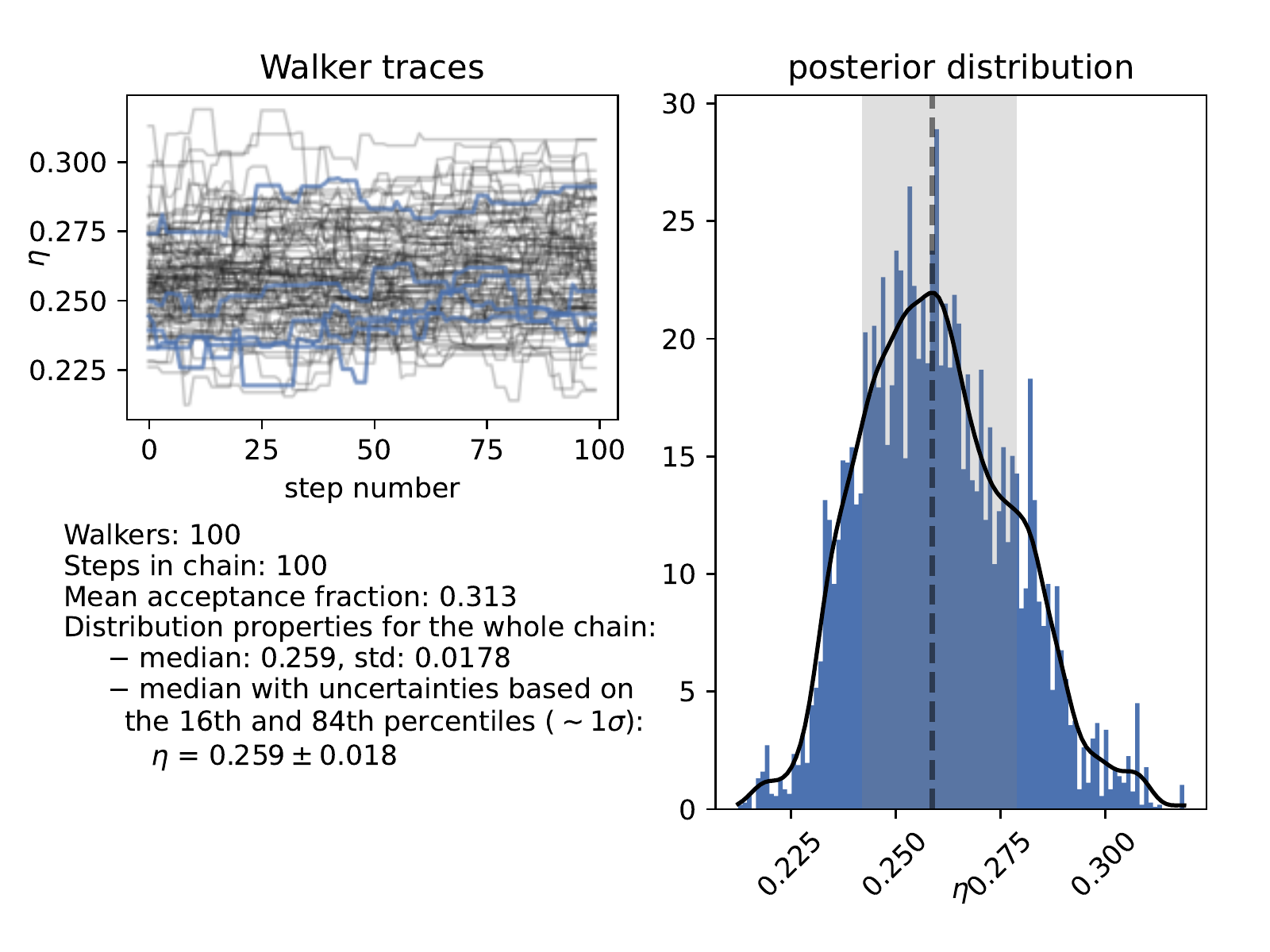}
\includegraphics[width=0.49\textwidth]{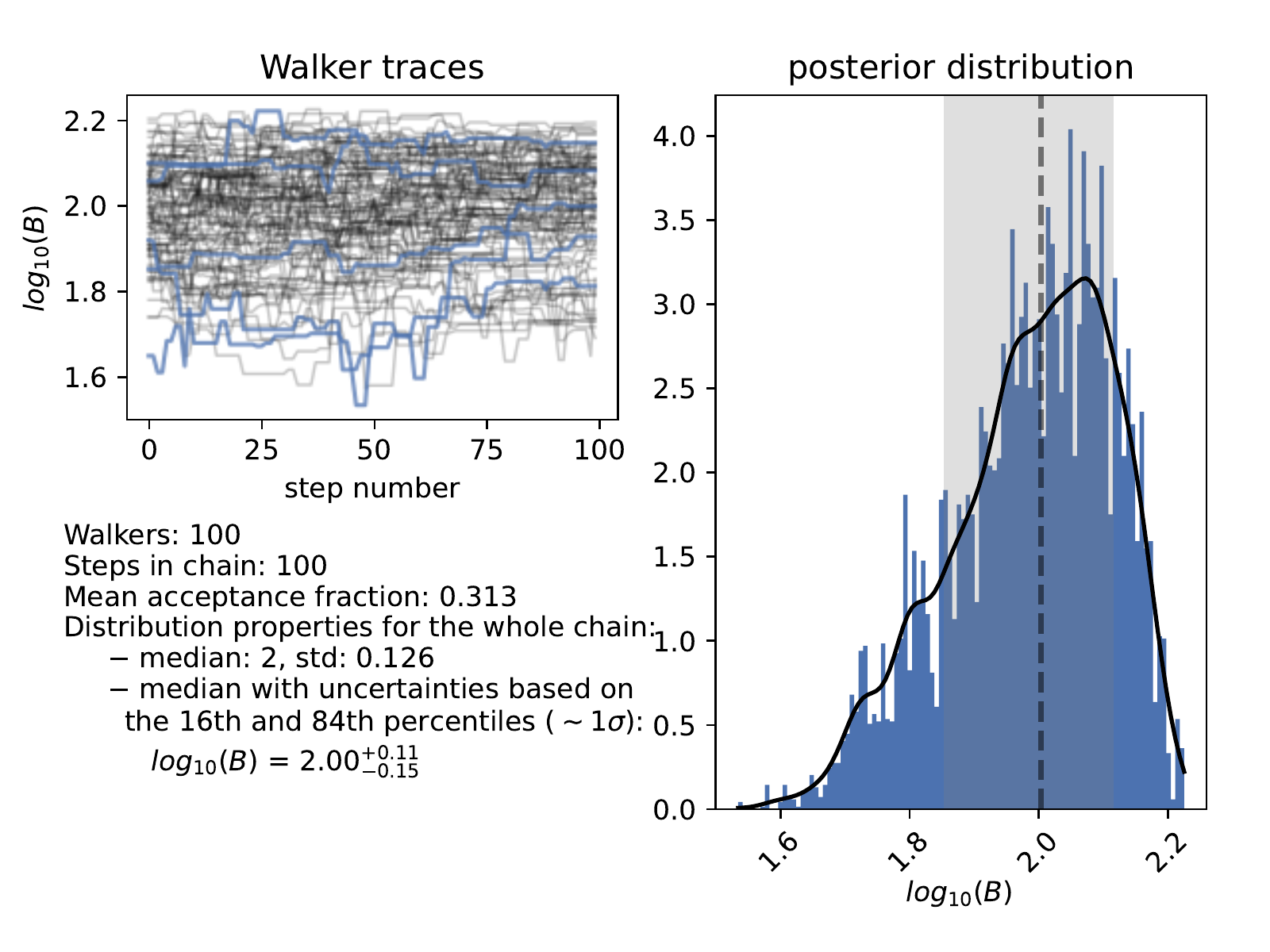}
\end{minipage}
\caption{The same as Fig.\ref{Figure6} but for the leptonic$-$hadronic hybrid model.}
\end{figure}
\setcounter{figure}{6}
\begin{figure}[h]
\centering
\begin{minipage}{\textwidth}
\includegraphics[width=0.49\textwidth]{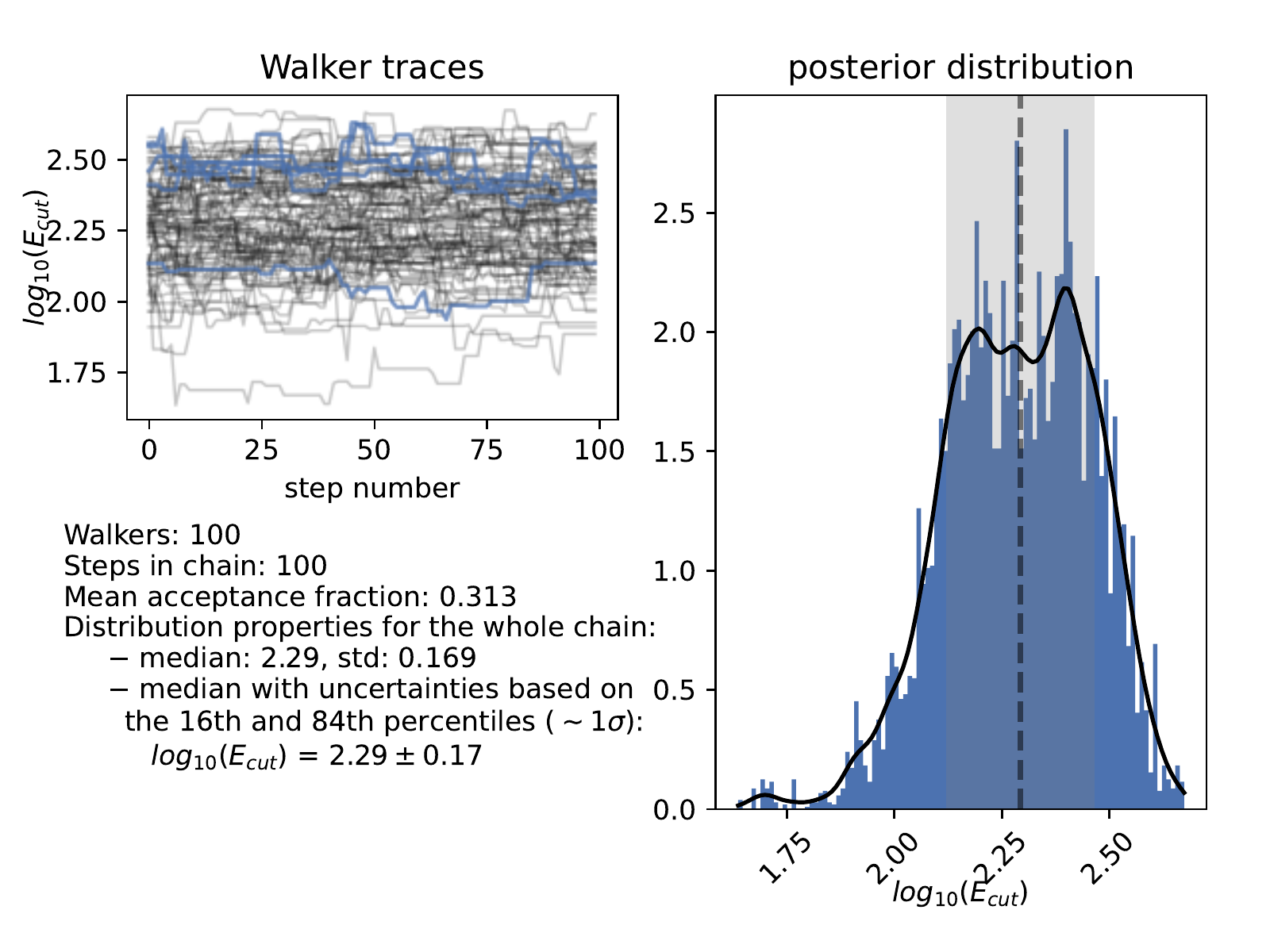}
\includegraphics[width=0.49\textwidth]{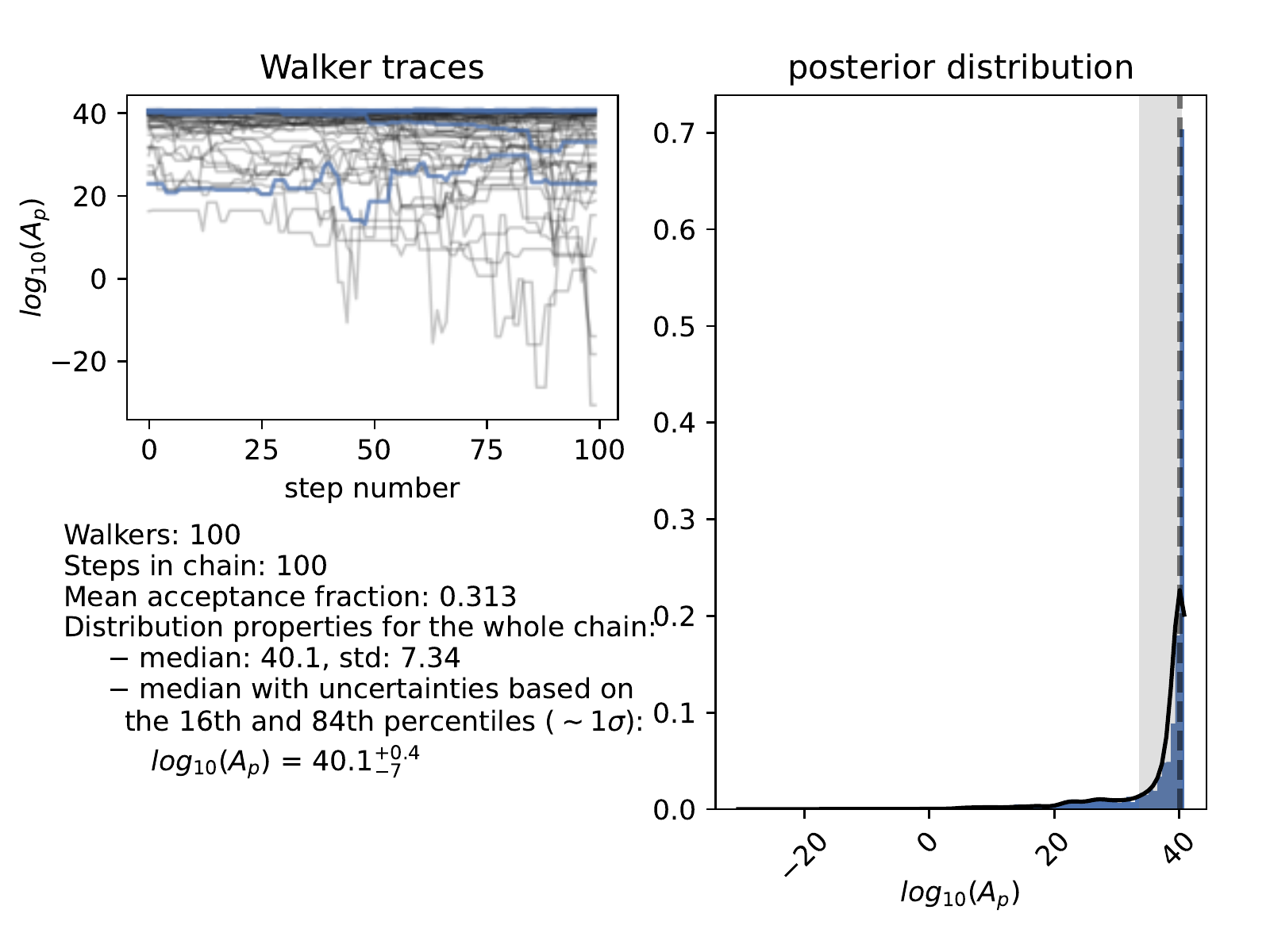}
\end{minipage}
\caption{(continued)}
\label{Figure7}
\end{figure}
\clearpage

\bibliography{crab}{}
\bibliographystyle{aasjournal}



\end{document}